\documentclass[journal, onecolumn, 12pt, draftcls]{IEEEtran}

\usepackage{mathrsfs}
\usepackage[noadjust]{cite}
\usepackage{graphicx,color,overpic,psfrag}
\usepackage{amsmath, amssymb}
\usepackage{latexsym}
\usepackage{bm}
\usepackage{amssymb}
\usepackage{cases}
\usepackage{array}
\usepackage{fancyhdr}
\usepackage{setspace}

\ifCLASSOPTIONcompsoc
\usepackage[caption=false,font=normalsize,labelfont=sf,textfont=sf]{subfig}
\else
\usepackage[caption=false,font=footnotesize]{subfig}
\fi

\usepackage{url}
\usepackage{algpseudocode}
\usepackage{algorithm}
\usepackage{blkarray}
\usepackage{booktabs}

\usepackage{multirow}
\usepackage{dsfont}
\usepackage{tabularx}
\usepackage[table]{xcolor}

\usepackage{amsfonts}

\usepackage{letltxmacro}

\graphicspath{{figure/}}

\newtheorem{lemma}{Lemma}

\newtheorem{prop}{Proposition}

\newcommand{\lmref}[1]{Lemma \ref{#1}}

\newcommand{\figref}[1]{Fig. \ref{#1}}

\newcommand{\alref}[1]{\textbf{Algorithm \ref{#1}}}
\newcommand{\appref}[1]{Appendix \ref{#1}}
\newcommand{\secref}[1]{Section \ref{#1}}

\newcommand{\propref}[1]{Proposition \ref{#1}}

\newcommand{\Exp}{{\mathsf{E}}}
\newcommand{\expect}[1]{\Exp\left\{#1\right\}}

\newcommand{\tr}[1]{\mathsf{tr}\left\{#1\right\}}
\newcommand{\diag}[1]{\mathsf{diag}\left\{#1\right\}}

\newcommand{\abs}[1]{\left|#1\right|}

\newcommand{\cK}{\mathcal{K}}

\newcommand{\cO}{\mathcal{O}}
\newcommand{\cP}{\mathcal{P}}

\newcommand{\cX}{\mathcal{X}}

\newcommand{\ba}{\mathbf{a}}

\newcommand{\bc}{\mathbf{c}}

\newcommand{\bn}{\mathbf{n}}

\newcommand{\bu}{\mathbf{u}}
\newcommand{\bv}{\mathbf{v}}

\newcommand{\bx}{\mathbf{x}}
\newcommand{\by}{\mathbf{y}}

\newcommand{\bA}{\mathbf{A}}
\newcommand{\bB}{\mathbf{B}}
\newcommand{\bC}{\mathbf{C}}
\newcommand{\bD}{\mathbf{D}}
\newcommand{\bE}{\mathbf{E}}
\newcommand{\bF}{\mathbf{F}}
\newcommand{\bG}{\mathbf{G}}
\newcommand{\bH}{\mathbf{H}}
\newcommand{\bI}{\mathbf{I}}

\newcommand{\bK}{\mathbf{K}}
\newcommand{\bL}{\mathbf{L}}

\newcommand{\bQ}{\mathbf{Q}}

\newcommand{\bS}{\mathbf{S}}

\newcommand{\bU}{\mathbf{U}}
\newcommand{\bV}{\mathbf{V}}
\newcommand{\bW}{\mathbf{W}}

\newcommand{\bzero}{\mathbf{0}}

\newcommand{\bLambda}{{\boldsymbol\Lambda}}
\newcommand{\balpha}{{\boldsymbol\alpha}}

\newcommand{\bGamma}{{\boldsymbol\Gamma}}
\newcommand{\bgamma}{{\boldsymbol\gamma}}
\newcommand{\bPsi}{{\boldsymbol\Psi}}
\newcommand{\bpsi}{{\boldsymbol\psi}}
\newcommand{\bPhi}{{\boldsymbol\Phi}}
\newcommand{\bphi}{{\boldsymbol\phi}}
\newcommand{\bOmega}{{\boldsymbol\Omega}}

\newcommand{\ntb}{\notag\\}

\newcommand{\R}{\mathbb{R}}
\newcommand{\C}{\mathbb{C}}

\newcommand{\I}{\mathbf{I}}

\newcommand{\rmG}{\mathrm{G}}
\newcommand{\rmR}{\mathrm{R}}

\newcommand{\Qk}{\mathbf{Q}_{k}}

\newcommand{\lambdak}{\mathbf{\Lambda}_{k}}

\newcommand{\Pmax}{P_{\mathrm{max}}}

\newcommand{\s}{\frac{1}{\sigma ^{2}}}

\newcommand{\tot}{\mathrm {tot}}

\newcommand{\BS}{\mathrm {BS}}
\newcommand{\AF}{\mathrm {r}}

\newcommand{\beamH}{\widetilde{\bH}}

\newcommand{\expectxn}[1]{\Exp_{\bx_{\mathrm{c}},\bn_{\mathrm{c}}}\left\{#1\right\}}
\newcommand{\Real}[1]{\Re \left\{#1\right\}}
\allowdisplaybreaks

\begin{document}

\title{ \LARGE Reconfigurable Intelligent Surfaces-Assisted Multiuser MIMO Uplink Transmission with Partial CSI}

\author{
Li~You, Jiayuan~Xiong, Yufei~Huang, Derrick~Wing~Kwan~Ng,
Cunhua~Pan,
Wenjin~Wang, and~Xiqi~Gao%
\thanks{This work will be presented in part at the IEEE International Conference on Communications, Dublin, Ireland, Jun. 2020 \cite{xiong2020Reconfigurable}.
}
\thanks{
L. You, J. Xiong, Y. Huang, W. Wang, and X. Q. Gao are with the National Mobile Communications Research Laboratory, Southeast University, Nanjing 210096, China, and also with the Purple
Mountain Laboratories, Nanjing 211100, China (e-mail: liyou@seu.edu.cn; jyxiong@seu.edu.cn; yufei\_huang@seu.edu.cn; wangwj@seu.edu.cn; xqgao@seu.edu.cn).
}
\thanks{
D. W. K. Ng is with the School of Electrical Engineering and Telecommunications, University of New South Wales, Sydney, NSW 2052, Australia (e-mail: w.k.ng@unsw.edu.au).
}
\thanks{
C. Pan is with the School of Electronic Engineering and Computer Science, Queen Mary University of London, London E1 4NS, U.K. (e-mail: c.pan@qmul.ac.uk).
}
}

\maketitle

\begin{abstract}
This paper considers the application of reconfigurable intelligent surfaces (RISs) (a.k.a. intelligent reflecting surfaces (IRSs)) to assist multiuser multiple-input multiple-output (MIMO) uplink transmission from several multi-antenna user terminals (UTs) to a multi-antenna base station (BS).
For reducing the signaling overhead, only partial channel state information (CSI), including the instantaneous CSI between the RIS and the BS as well as the slowly varying statistical CSI between the UTs and the RIS, is exploited in our investigation.
In particular, an optimization framework is proposed for jointly designing the transmit covariance matrices of the UTs and the RIS phase shift matrix to maximize the system global energy efficiency (GEE) with partial CSI.
We first obtain closed-form solutions for the eigenvectors of the optimal transmit covariance matrices of the UTs. Then, to facilitate the design of the transmit power allocation matrices and the RIS phase shifts, we derive an asymptotically deterministic equivalent of the objective function with the aid of random matrix theory. We further propose a suboptimal algorithm to tackle the GEE maximization problem with guaranteed convergence, capitalizing on the approaches of alternating optimization, fractional programming, and sequential optimization. Numerical results substantiate the effectiveness of the proposed approach as well as the considerable GEE gains provided by the RIS-assisted transmission scheme over the traditional baselines.
\end{abstract}



\section{Introduction}
The immense demand for delivering high-quality wireless communication services continues to grow and will be never-ending, especially in the fifth-generation (5G) and beyond era \cite{Zhang2019Multiple}. Those ubiquitous communication services have numerous exceptionally high requirements, such as ultra-low latencies, excellent spectral efficiency (SE), reliability, wireless charging, and high energy efficiency (EE), and therefore pose new challenges in 5G and beyond wireless networks \cite{Wong2017key}. To cope with these challenges, more and more radically new technologies emerge for future wireless communications. Among these advanced approaches, a brand-new research direction, named reconfigurable intelligent surface (RIS) (also known as intelligent reflecting surface (IRS)), has recently received tremendous interests and is promising to pave its way to the mainstream of wireless communications \cite{Zhang2019Multiple}.

RISs are artificial programmable surfaces of electromagnetic materials and their reflection properties can be controlled by integrated electronics \cite{basar2019wireless}. As a two-dimensional application of meta-materials \cite{cui2014coding,liaskos2018new}, this promising new hardware technology owns several distinctive characteristics and remarkable potentials. For instance, RISs can create a so-called smart radio environment \cite{di2019smart,qingqing2019towards}. In practice, the smart radio environment is generally a wireless network, where the propagation environment is controllable and reconfigurable. In other words, the application of RISs can establish a favourable communication channel which facilitates signal processing and information transmission. As a result, destructive effects caused by multi-path components and Doppler shifts in random and uncontrollable surroundings can be alleviated and system performances are enhanced. In addition, the full-band response makes RISs appealing in numerous practical applications such as millimeter-wave and Terahertz communications \cite{basar2019wireless}. RISs are also environmentally friendly because their reflecting elements are almost passive and ideally require no dedicated power sources. Moreover, thermal noises at receivers have no impact on RISs as active power amplifiers are not necessary. Consequently, noises would not be introduced or magnified during signal reflection. Meanwhile, low hardware footprints of RIS structures allow high-flexibility and low-implementation cost to install RISs, i.e., on factory ceilings, rooms, building facades, and even onto human clothing. Due to the aforementioned appealing properties, RIS is significantly different from the existing technologies, e.g., relay and backscatter communications \cite{huang2019reconfigurable,wu2019intelligent}.

The distinguishable features of RISs not only enable an emerging RIS-empowered environment, but also introduce a paradigm shift in wireless transmission designs and optimizations. The applications of meta-surfaces have been extensively researched in, e.g., radar and satellite communications. Recently, attentions have been paid to focusing on terrestrial mobile communications with RISs \cite{shen2019secrecy,cui2019secure,yu2019enabling,yang2019intelligent,wu2019intelligent,pan2019intelligent,huang2019reconfigurable}. For example, contributions in \cite{shen2019secrecy,cui2019secure,yu2019enabling} exploited the use of RISs to enhance the physical layer security, considering the case with only one legitimate receiver and one eavesdropper. The single-user wireless transmission was studied in \cite{yang2019intelligent,wu2019intelligent} with different objectives, such as maximizing the system data rate \cite{yang2019intelligent} or received signal power \cite{wu2019intelligent}. In addition, RIS-assisted multiuser multiple-input single-output downlink systems were investigated in various literatures such as \cite{huang2019reconfigurable,pan2019intelligent}. Due to the significant potentials offered by the RISs to enhance the performance of wireless networks, plenty of related contributions have appeared recently, e.g., the survey papers in \cite{basar2019wireless,cui2014coding,liaskos2018new,di2019smart,qingqing2019towards} and references therein.

The performance of RIS-assisted wireless transmission highly depends on the adaptivity of the RIS elements. Generally, the RIS phase parameters are adapted to the channel states for improving the system performance. One of the major design concerns is how quickly the RIS phases can be tuned in practice.
For fixed or low-mobility transmission scenarios, the channel states vary over time slowly, and thus it is possible to perform RIS phase tuning exploiting full instantaneous channel state information (CSI). It is worth remarking that most of the existing resource allocation strategies for RIS-assisted wireless networks were carried out by assuming the availability of perfect knowledge of full CSI, including the instantaneous CSI between the RIS and the base station (BS) or the user terminals (UTs), e.g., \cite{yu2019enabling,cui2019secure,shen2019secrecy,yang2019intelligent,wu2019intelligent,pan2019intelligent,huang2019reconfigurable}. However, in high mobility scenarios with fast time-varying channels, tuning RIS parameters (as well as transmit precoding) via exploiting instantaneous CSI is challenging due to the following reasons. First, in a short duration of coherence time, system resources have to be reallocated and the RIS phase shift parameters have to be updated frequently, thus incurring significant signaling overhead \cite{Zappone2020Overhead}. Second, RISs are usually equipped with smart controllers which adapt biasing voltages according to the available CSI for realizing phase tuning in practice \cite{qingqing2019towards}. Although the RIS itself operates without consuming any additional transmit power ideally, the smart controller will still be power-consuming if it is overloaded with continuous operations, i.e., frequently tuning the RIS elements would not be energy efficient. Therefore, it is natural to exploit the slowly varying channel properties for resource allocation in RIS-assisted wireless networks.
Indeed, devising resource allocation strategies by capitalizing on partial CSI such as the slowly time-varying statistical CSI, which varies a much longer time scale compared to instantaneous CSI, has become an important research trend in various scenarios, e.g., \cite{You15Pilot,Gao09Statistical,You16Channel,Wen11On,You17BDMA,Zappone2014energye,You2020energy,Lu16Free,You20LEO,Tulino06Capacity,You2020Pilot}.

Given the above considerations, we investigate energy-efficient transmit precoding and RIS tuning strategies for RIS-assisted multiuser multiple-input multiple-output (MIMO) uplink transmission from several multi-antenna UTs to the multi-antenna BS with the consideration of partial CSI. Specifically, since the positions of the RIS and the BS are both fixed, the RIS-to-BS channel is slowly time-varying and its instantaneous CSI can naturally be acquired. In contrast, for the UT-to-RIS channels, it is reasonable to exploit its statistical CSI for resource allocation as UTs are usually in mobility, leading to fast time-varying channels.
Recently, there have been some initial attempts investigating RIS-assisted transmission applying statistical CSI, e.g., \cite{Han19Surface,Zhang19surface,Zhao19Intelligent}. In \cite{Han19Surface}, the performance of the RIS-assisted large-scale antenna system exploiting statistical CSI was investigated. In \cite{Zhang19surface}, the outage probability of the RIS-assisted system adopting statistical CSI was investigated. In \cite{Zhao19Intelligent}, a two-timescale beamforming approach was proposed for RIS enhanced transmission.
Note that these works all focused on the cases where UTs are quipped with only one antenna. However, current standards have advocated multi-antenna UTs to improve the transmission performance \cite{Zhang2019Multiple,Wong2017key}.
To the authors' best knowledge, this is the first work that exploits statistical CSI for resource allocation in RIS-assisted multiuser MIMO uplink networks involving several multi-antenna UTs. The problem is challenging to handle due to the complicated problem structure and existing approaches for single-antenna UTs cannot be applied to our investigations.
The main contributions of this paper are summarized as follows:
\begin{itemize}
\item We investigate resource allocation strategy design in RIS-assisted multiuser MIMO uplink systems with partial CSI. Considering the global energy efficiency (GEE) as the design objective, we formulate the optimization problem to jointly design the transmit covariance matrices of all UTs and the phase shifts of the RIS elements subject to the maximum transmit power constraint at each UT. The alternating optimization (AO) method is adopted to address this sophisticated GEE maximization problem, which facilitates the design of a computationally efficient iterative resource allocation algorithm.
\item To optimize the transmit covariance matrices of all UTs with a fixed RIS phase shift matrix, we first obtain the optimal transmit signal directions at the UT sides in a closed-form. We further derive an asymptotically deterministic equivalent (DE) of the objective function to simplify the problem. Later, the concave-convex fractional power allocation problem is tackled by applying Dinkelbach's algorithm.
\item To handle the challenging RIS phase shift matrix optimization problem, we introduce an equivalent mean-square error (MSE) minimization problem. Utilizing the inherent structure of the MSE minimization problem, we develop a novel approach to optimize the RIS phase shift values, based on the block coordinate descent (BCD) method and the minorization-maximization (MM) technique.
\item Uniting all the methods adopted above, we present a well-structured and low-complexity algorithm with guaranteed convergence for GEE maximization (as well as SE maximization) in the RIS-assisted multiuser MIMO uplink transmission. Numerical simulations are conducted to validate the potentials of exploiting RISs for promoting system performances. The results verify the capability of the proposed approach to obtain higher GEE performance compared to that of the conventional baselines.
\end{itemize}

The rest of this paper is organized as follows. In \secref{sec:sysmod}, we describe the channel model of the considered RIS-assisted multiuser MIMO uplink system and formulate the corresponding GEE maximization problem. In \secref{sec:Optimization_Q}, the transmit covariance matrices at the UT sides are optimized with the closed-form solutions of the eigenvectors and a Dinkelbach-based power allocation algorithm. In \secref{sec:Optimization_Phi}, the optimization of the RIS phase shift matrix is performed by handling an equivalent MSE minimization problem through a BCD-based algorithm. In \secref{sec:overallopt}, we combine the methods adopted in the previous two sections and then present an overall approach for addressing the resource allocation problem of RIS-assisted MIMO uplink transmission with partial CSI. The numerical results are provided in \secref{sec:numerical_results}. The conclusion is drawn in \secref{sec:conclusion}.

\emph{Notations:} $\Real{\cdot}$ represents the real part of a complex value. Matrices and column vectors are denoted by upper and lower case boldface letters, respectively. $\bI_{N}$ denotes an identity matrix with subscript $N$ being the matrix dimension. The operators $\expect{\cdot}$, $\tr{\cdot}$, and $\det(\cdot) $ represent the expectation, trace, and determinant operations, respectively. The superscripts $(\cdot)^{-1}$, $(\cdot)^T$, and $(\cdot)^H$ are denoted as the inverse, transpose, and conjugate-transpose operations, respectively. $\mathcal{CN}(\ba,\bB)$ represents the circular symmetric complex-valued Gaussian distribution with mean $\ba$ and covariance matrix $\bB$. The inequality $\bA \succeq \bzero$ indicates that $\bA$ is a positive semi-definite matrix. $\odot$ denotes the Hadamard product. The operator $\diag{\bx}$ generates a diagonal matrix with the elements of $\bx$ along its main diagonal. $\| \bx \|$ denotes the Euclidean norm of $\bx$. The notation $\triangleq$ is utilized for definitions and $\jmath = \sqrt{-1}$ denotes the imaginary unit.

\section{System Model}\label{sec:sysmod}
This section first introduces the channel model of the considered RIS-assisted multiuser MIMO uplink system and then describes the energy consumption model of the system. In addition, the joint design of the RIS phase shifts and the transmit covariance matrices at the UT sides with partial CSI is formulated as an optimization problem in the last part of this section.

\subsection{Channel Model}
\begin{figure}[!t]
\centering
\includegraphics[width=0.7\textwidth]{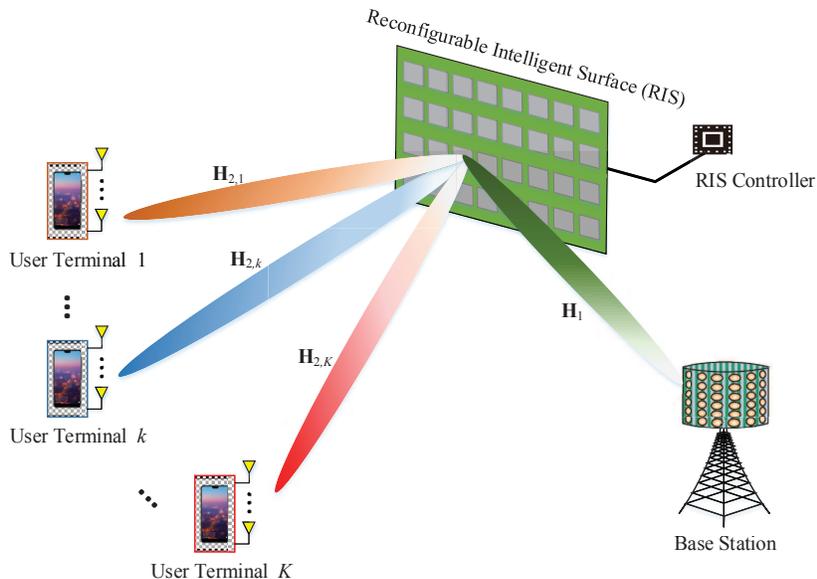}
\caption{The considered RIS-assisted multiuser MIMO uplink system.}
\label{fig:RIS}
\end{figure}

The considered RIS-assisted multiuser MIMO uplink communication system is sketched in \figref{fig:RIS}, consisting of $K$ UTs, one RIS, and one BS. We assume that each UT $k \in {\cal K} \triangleq \left\{ {1,2, \ldots ,K} \right\}$ is equipped with $N_k$ transmit antennas to convey signals while the BS has $M$ antennas for receiving. The communication is enhanced via the deployment of a RIS composed of $N_{\rmR}$ reflecting units, which is capable of applying phase shifts reacting to the incoming signals. Due to the unfavorable propagation conditions, the direct UT-to-BS channel is negligible and therefore ignored in the system model, as commonly adopted in the literature \cite{huang2019reconfigurable}. In addition, the signals reflected by the RIS more than once are also ignored due to, e.g., high path loss \cite{pan2019multicell}.

We denote $\bx_k \in \C^{N_k \times 1} $ as the signal vector sent by UT $k$, which satisfies $ \expect { \bx_k}  = \bzero$ and $ \expect { \bx_k \bx_{k'}^H}  = \bzero$, $\forall k' \ne k$. The covariance matrix of transmit signal $\bx_k$ at UT $k$ is denoted by $\bQ_k = \expect{\bx_k \bx_k^H} \in \C^{N_k \times N_k}$. Then, the received signal at the BS is
\begin{align}
\by = \sum\limits_{k=1}^K { \bH_1 \bPhi \bH_{2,k} \bx_k } + \bn,
\end{align}
where $\bH_1 \in \C^{M \times N_{\rmR}} $ represents the channel matrix from the RIS to the BS, $\bH_{2,k} \in \C^{N_{\rmR} \times N_k} $ denotes the channel matrix between UT $k$ and the RIS with its $(n,m)$th entry being the complex-valued channel coefficient from the $m$th antenna at UT $k$ to the $n$th element of the RIS, and $\bn \sim \mathcal{CN}(\bzero,\sigma^2 \bI_{M})$ is the thermal noise at the BS with $\sigma^2$ being the noise power. In addition, we adopt an ideal RIS model where only the phases of the incoming signals are adjustable while the amplitudes are kept constant \cite{huang2019reconfigurable,wu2019intelligent}. Then, the operation of the RIS is described by the diagonal matrix $\bPhi = \diag{\phi_1,\ldots,\phi_{N_{\rmR}}}$, where $\phi_n = \mathrm{e}^{\jmath \theta_n},n\in\left\{1,\ldots,N_{\rmR}\right\}$, with $\theta_n$ being the phase shift introduced by the $n$th element of the RIS.

In this work, we consider the jointly spatially correlated Rayleigh fading channel model \cite{Gao09Statistical}, and the UT-to-RIS channel $\bH_{2,k}$ exhibits the structure as
\begin{align}\label{eq:beam_H2}
\bH_{2,k} & = \bU_{2,k} \beamH_{2,k} \bV^H_{2,k},\quad \forall k,
\end{align}
where $\bU_{2,k} = \left[ \bu_{2,k,1}, \bu_{2,k,2}, \ldots, \bu_{2,k,N_{\rmR}} \right] \in \C^{N_{\rmR} \times N_{\rmR}}$ and $\bV_{2,k} = \left[ \bv_{2,k,1}, \bv_{2,k,2}, \ldots, \bv_{2,k,N_{k}} \right] \in \C^{N_k \times N_k}$ are both deterministic and unitary matrices, and the ${N_{\rmR} \times N_{k}}$ complex-valued matrix $\beamH_{2,k}$ is random with elements being zero-mean and independently distributed. In addition, the statistics of UT-to-RIS channel $\beamH_{2,k}$ is given by
\begin{align}
\bOmega_k = \expect { \beamH_{2,k} \odot \beamH_{2,k}^* }  \in { \R ^{ N_{\rmR} \times N_k }},
\end{align}
with its $(n,m)$th element specifying the average energy coupled between $\bu_{2,k,n}$ and $\bv_{2,k,m}$ \cite{Gao09Statistical}. Accordingly, $\bOmega_k$ is known as the eigenmode coupling matrix of the channel between UT $k$ and the RIS. Note that the channel statistics, $\bOmega_k$, $\forall k$, rather than the instantaneous channel realizations, $\bH_{2,k}$, $\forall k$, will be exploited in our resource allocation design. In addition, the instantaneous knowledge of the RIS-to-BS channel $\bH_1$ is assumed to be fully known as the locations of the RIS and the BS are usually fixed. Then, the ergodic SE of the RIS-assisted multiuser MIMO uplink system is given by \cite{Wen11On}
\begin{align}\label{eq:ergodic_rate}
R =  \expect { \log_2 \det  \left( \bI_M + \s \sum\limits_{k=1}^K \bH_1 \bPhi \bH_{2,k} \bQ_k \bH^H_{2,k}  \bPhi^H \bH^H_1 \right) } \ [\mathrm{bits/s/Hz}],
\end{align}
where the expectation is taken with respect to $\bH_{2,k}$, $\forall k$.

\subsection{Energy Consumption Model}
The total energy consumption of our considered RIS-assisted multiuser MIMO uplink system is constituted by three major parts, including the transmit power, the hardware static power, and the RIS power consumption.
We adopt a general affine model for the power consumption. In particular, the total amount of the energy consumption of our considered RIS-assisted multiuser MIMO uplink system is given by \cite{huang2019reconfigurable}
\begin{align}\label{eq:total_power_consumption}
P_{\tot} = \sum\limits_{k=1}^K \left( \xi_k \tr{\bQ_k} + P_{\mathrm{c},k} \right) + P_{\mathrm{BS}} + N_{\rmR} P_{\mathrm{s}}.
\end{align}
In \eqref{eq:total_power_consumption}, $\xi_k = \rho_k^{-1}$ with $\rho_k$ denoting the transmit power amplifier efficiency at UT $k$, $\tr{\bQ_k}$ denotes the average transmit power consumed by UT $k$, and $P_{\mathrm{c},k}$ represents the static circuit power dissipation at UT $k$. In addition, $P_{\mathrm{BS}}$ and $N_{\rmR}P_{\mathrm{s}}$ incorporate the static hardware-dissipated power at the BS and the RIS, respectively.\footnote{Note that the affine power consumption model requires that the following assumptions are satisfied. First, the static circuit power $P_{\mathrm{c},k}$ is independent of the data rate. Second, the power amplifiers at all UTs operate in the linear region of their corresponding transfer functions, where a constant power offset can well approximate the hardware-consumed power. Typical wireless communication transceivers satisfy these two assumptions generally \cite{huang2019reconfigurable}.} It is worth emphasizing that there is no transmit power consumed by the RIS because the reflectors of the RIS are passive elements which do not change the magnitude of the reflected signals. In fact, the potential amplification gain offered by the RIS is realized via appropriately adjusting the phase shifts of the reflectors so that the impinging signals are coherently combined at the desired receiver.

\subsection{Problem Formulation}
In this work, we investigate the transmission strategy for the RIS-assisted multiuser MIMO uplink system, where we strive to jointly optimize the transmit covariance matrices, $\bQ_k$, $\forall k$, at the UT sides and the diagonal RIS phase shift matrix, $\bPhi$, for improving the system performance. From a systematic perspective, we adopt the GEE of the entire communication system as our design criterion. From \eqref{eq:ergodic_rate} and \eqref{eq:total_power_consumption}, we define the system GEE as
\begin{align}\label{eq:geedef}
\mathrm{GEE}\left( \bQ,\bPhi \right) \triangleq W \frac{ \expect { \log_2 \det  \left( \bI_M + \s \sum\limits_{k=1}^K \bH_1 \bPhi \bH_{2,k} \bQ_k \bH^H_{2,k}  \bPhi^H \bH^H_1 \right) }   }{ \sum\limits_{k=1}^K \left( \xi_k \tr{\bQ_k} + P_{\mathrm{c},k} \right) + P_{\mathrm{BS}} + N_{\rmR} P_{\mathrm{s}} } \ [\mathrm{bits/Joule}],
\end{align}
where $W$ is the system bandwidth. Then, the optimization of $\bQ_k$, $\forall k$, and $\bPhi$ is formulated as the following problem
\begin{subequations}\label{eq:problem_Q_phi}
\begin{align}
\cP_{\bQ,\bPhi}:\quad\underset{\bQ,\bPhi} \max \quad & \mathrm{GEE}\left( \bQ,\bPhi \right)\\
{\mathrm{s.t.}}\quad
& \tr { \bQ_k } \le P_{\max,k}, \quad \bQ_k \succeq \bzero,\quad\forall k \in \cK, \\ \label{eq:phi_constraint}
& \left| \phi_n \right| = 1,\quad n = 1,\ldots,N_{\rmR},
\end{align}
\end{subequations}
where $\bQ \triangleq \left\{ \bQ_k \right\}_{k=1}^K$ and $P_{\max,k}$ is the maximum available transmit power at UT $k$. The constraints in \eqref{eq:phi_constraint} ensure that the RIS reflectors only operate as phase shifters which do not provide any amplification gain to the incoming signals. Notice that if we set $\xi_k = 0$ for all UTs, the denominator of the objective function in \eqref{eq:problem_Q_phi} becomes independent of both $\bQ$ and $\bPhi$, thus is regarded as a constant. Consequently, the fractional objective function in $\cP_{\bQ,\bPhi}$ is reduced into a non-fractional form where only the numerator, i.e., the system SE, has to be maximized. Hence, problem $\cP_{\bQ,\bPhi}$ can be utilized to investigate not only the GEE maximization, but also the SE maximization in the RIS-assisted MIMO uplink transmission with partial CSI.

It is worth noting that the optimization problem $\cP_{\bQ,\bPhi}$ in \eqref{eq:problem_Q_phi} is quite challenging due to the following reasons. First, computing the objective function of $\cP_{\bQ,\bPhi}$ with expectation operations is computationally expensive. Second, the fractional objective function in \eqref{eq:geedef} makes $\cP_{\bQ,\bPhi}$ essentially an NP-hard problem \cite{zappone2015energy}. Additionally, the presence of the unit-modulus-constrained $\bPhi$ further complicates the optimization procedure. In the following, we aim to develop an efficient approach to address this difficult problem.

\section{Optimization of Transmit Covariance Matrices}\label{sec:Optimization_Q}
Note that problem $\cP_{\bQ,\bPhi}$ involves two matrix variables, $\bQ$ and $\bPhi$, which are complicated to be jointly optimized. To tackle $\cP_{\bQ,\bPhi}$ more conveniently, we resort to AO, which is applicable to optimization problems with different blocks of variables. In particular, we solve for $\bQ$ and $\bPhi$ iteratively, i.e., optimize $\bQ$ with a fixed $\bPhi$ and optimize $\bPhi$ with a fixed $\bQ$.

Following the principle of AO, we first consider the design of the transmit covariance matrices $\bQ_k$ of all UTs with an arbitrarily given $\bPhi$, which is characterized as
\begin{align}\label{eq:problem_Q}
\cP_{\bQ}:\quad\underset{\bQ} \max \quad & \frac{ \expect { \log_2 \det  \left( \bI_M + \s \sum\limits_{k=1}^K \bH_1 \bPhi \bH_{2,k} \bQ_k \bH^H_{2,k}  \bPhi^H \bH^H_1 \right) }   }{ \sum\limits_{k=1}^K \left( \xi_k \tr{\bQ_k} + P_{\mathrm{c},k} \right) + P_{\mathrm{BS}} + N_{\rmR} P_{\mathrm{s}} } \ntb
{\mathrm{s.t.}}\quad
& \tr { \bQ_k } \le P_{\max,k}, \quad \bQ_k \succeq \bzero,\quad\forall k \in \cK.
\end{align}
Notice that the bandwidth $W$ is omitted in the above problem as well as in the sequel since it is a constant and does not have any impact on the optimization. The number of variables in $\cP_{\bQ}$ is smaller than that in $\cP_{\bQ,\bPhi}$. More importantly, the non-convex constraints of $\cP_{\bQ,\bPhi}$ in \eqref{eq:phi_constraint} are independent of $\bQ$ and are ignored. Hence, $\cP_{\bQ}$ is relatively easier to manage when compared with $\cP_{\bQ,\bPhi}$. However, it is still inconvenient to deal with $\cP_{\bQ}$ due to the relatively large number of variables. To this end, we first provide the eigenvalue decomposition of $\bQ_k$ as
\begin{align}\label{eq:EVD_Q}
\bQ_k & = \bV_{k} \bLambda_k \bV_{k}^H,
\end{align}
which further decomposes $\bQ_k$ into two blocks, the unitary eigenmatrix $\bV_{k} \in \C^{N_k \times N_k}$ and the diagonal power allocation matrix $\bLambda_k \in \R^{N_k \times N_k}$. In fact, $\bV_k$ and $\bLambda_{k}$ specify the transmit subspace of UT $k$, composed of signal directions and the power distributed in each dimension of the subspace, respectively. With a slight abuse of notations, we define $\bLambda \triangleq \left\{ \bLambda_k \right\}_{k=1}^K$ and $\bV \triangleq \left\{ \bV_k \right\}_{k=1}^K$ for later use. Then, according to the decoupling of these two variables, we perform AO again to optimize $\bQ$ by alternatingly solving for $\bV$ and $\bLambda$.

\subsection{Optimal Transmit Directions at UTs}
We begin with finding the optimal transmit signal directions for all UTs. Specifically, the design of $\bV$ with both $\bPhi$ and $\bLambda$ fixed is characterized in the following optimization problem
\begin{align}\label{eq:subproblem_direction}
\cP_{\bV}:\quad\underset{ \bV}  \max \quad & \frac{ \expect { \log_2 \det  \left( \bI_M + \s \sum\limits_{k=1}^K \bH_1 \bPhi \bH_{2,k} \bV_{k} \bLambda_k \bV_{k}^H \bH^H_{2,k}  \bPhi^H \bH^H_1 \right) }   }{ \sum\limits_{k=1}^K \left( \xi_k \tr{\bV_{k} \bLambda_k \bV_{k}^H} + P_{\mathrm{c},k} \right) + P_{\mathrm{BS}} + N_{\rmR} P_{\mathrm{s}} } \ntb
{\mathrm{s.t.}}\quad
& \bV_{k} \bV_{k}^H = \I_{N_k} , \quad \forall k.
\end{align}
The optimal solution to the above problem is presented in the following proposition. The proof is similar as that in e.g., \cite{Tulino06Capacity,You2020energy} and therefore omitted for brevity.
\begin{prop}\label{theorem:beam_domain_optimal}
For arbitrarily given $\bPhi$ and $\bLambda$, the eigenmatrix of the optimal $\Qk$ for any UT $k$ is identical with the corresponding $\bV_{2,k}$, which appears in \eqref{eq:beam_H2}, i.e.,
\begin{align}
\bV_{k} = \bV_{2,k}, \quad \forall k.
\end{align}
\end{prop}

Applying \emph{\propref{theorem:beam_domain_optimal}}, in order to achieve the maximum GEE of the RIS-assisted uplink system, the optimal transmit subspace of each input $\bx_k$ at UT $k$ should be in the signal space spanned by the eigenmatrix of the corresponding channel's transmit correlation matrix.

Note that \emph{\propref{theorem:beam_domain_optimal}} holds for arbitrarily given $\bPhi$ or $\bLambda$. In other words, the two-layer AO, i.e., one layer as the iterative optimization between $\bQ$ and $\bPhi$, and the other between $\bV$ and $\bLambda$, is boiled down to one layer where we only need to alternatingly optimize $\bLambda$ and $\bPhi$. In particular, by setting $\bV_{k} = \bV_{2,k}$, $\forall k$, we can directly consider the optimization problem with respect to $\bLambda$, which is given by
\begin{align}\label{eq:power_allocation}
\cP_{\bLambda}:\quad \underset{\bLambda} \max \quad & \frac{ \expect { \log_2 \det  \left( \bI_M + \s \sum\limits_{k=1}^K \bH_1 \bPhi \bU_{2,k} \beamH_{2,k} \bLambda_k \beamH^H_{2,k}  \bU^H_{2,k} \bPhi^H \bH^H_1 \right) }   }{ \sum\limits_{k=1}^K \left( \xi_k \tr{\bLambda_k} + P_{\mathrm{c},k} \right) + P_{\mathrm{BS}} + N_{\rmR} P_{\mathrm{s}} } \ntb
{\mathrm{s.t.}}\quad
& \tr { \bLambda_k }  \le P_{\max,k}, \quad \lambdak \succeq \bzero,\quad \lambdak\; \mathrm{diagonal},\quad \forall k\in {\cal K}.
\end{align}

\subsection{DE Method}
Computing the numerator while dealing with problem $\cP_{\bLambda}$ is generally resource-consuming as it requires to compute the expectation values with respect to the UT-to-RIS channels $\bH_{2,k}$ for all UTs which involve high-dimensional integrals. The traditional Monte Carlo method for calculating the expectations via channel averaging is also computationally expensive. Instead, by leveraging random matrix theory \cite{Couillet11Random,Lu16Free}, we derive deterministic and asymptotically tight approximations of the expectations needed by the numerator of the objective function in $\cP_{\bLambda}$ when the numbers of antennas $M$ and $N$ both tend to infinity but with a constant ratio. For notational simplicity, we define $\bD = \diag{\bLambda_1,\bLambda_2,\cdots,\bLambda_K} \in \R^{N \times N}$ and $\bG = \left[ \bG_1 \ \bG_2 \cdots \bG_K  \right] \in \C^{ M \times N}$ where $\bG_k = \bH_1 \bPhi \bU_{2,k} \beamH_{2,k} \in \C^{ M \times N_k}$, $\forall k$, and $N = \sum\nolimits_k{N_k}$. Then, the numerator of the objective function in \eqref{eq:power_allocation} is recast as the following compact form
\begin{align}\label{eq:R_bigLambda}
R\left(\bLambda\right) & = \expect { \log_2 \det  \left( \bI_M + \s \bG \bD \bG^H \right) }.
\end{align}
With this reformulation, we take advantage of the existing results in \cite{Wen11On} and obtain the DE of $R\left(\bLambda\right)$ in \eqref{eq:R_bigLambda} as
\begin{align}\label{eq:DE}
\overline{R} \left( \bLambda \right) = \sum\limits_{k=1}^K \log_2 \det \left( \bI_{N_k} + \bGamma_k \bLambda_k \right) + \log_2 \det \left( \bI_M + \bPsi \right) - \sum\limits_{k=1}^K {\bgamma_k^T \bOmega_k \bpsi_k},
\end{align}
where $\bgamma_k \triangleq \left[ \gamma_{k,1},\gamma_{k,2},\ldots,\gamma_{k,N_{\rmR}} \right]^T$, $\bpsi_k \triangleq \left[ \psi_{k,1},\psi_{k,2},\ldots,\psi_{k,N_k} \right]^T$, and $\bPsi \triangleq \sum\nolimits_k{\bPsi_k} \in \C^{M \times M}$. In addition, defining $\bU_{\rmG_k} \triangleq \bH_1 \bPhi \bU_{2,k} \in \C^{ M \times N_{\rmR}}$, $\forall k$, we calculate
\begin{align}\label{eq:T}
\bGamma_k & =  \diag{ \bOmega_k^T \bgamma_k } \in \C^{N_k \times N_k}, \\ \label{eq:F}
\bPsi_k & = \s \bH_1 \bPhi \bU_{2,k} \diag{ \bOmega_k \bpsi_k } \bU_{2,k}^H \bPhi^H \bH_1^H
 = \s \bU_{\rmG_k} \diag{ \bOmega_k \bpsi_k } \bU_{\rmG_k}^H \in \C^{ M \times M }.
\end{align}
Moreover, the quantities $\bgamma \triangleq \left\{\gamma_{k,m}\right\}_{\forall k,m}$ and $\bpsi \triangleq \left\{\psi_{k,n}\right\}_{\forall k,n}$ are calculated as the unique solutions to the following iterative equations:
\begin{align}\label{eq:gamma}
\gamma_{k,m} & = \s \bu_{\rmG_k,m}^H \left( \bI_M + \bPsi \right)^{-1} \bu_{\rmG_k,m}, \quad k = 1,\ldots,K, \ m = 1,\ldots,N_{\rmR}, \\ \label{eq:psi}
\psi_{k,n} 
& = \left[ \bLambda_k \left( \bI_{N_k} + \bGamma_k \bLambda_k \right)^{-1} \right]_{n,n} \ntb
& = \frac{\lambda_{k,n,n}}{1+g_{k,n,n} \lambda_{k,n,n}}, \quad \quad \quad \quad \quad k = 1,\ldots,K, \ n = 1,\ldots,N_k,
\end{align}
where $\bu_{\rmG_k,m}$ is the $m$th column of $\bU_{\rmG_k}$, i,e., $\bU_{\rmG_k} = \left[ \bu_{\rmG_k,1},\bu_{\rmG_k,2},\ldots,\bu_{\rmG_k,N_{\rmR}} \right]$. Lastly, $\lambda_{k,n,n}$ and $g_{k,n,n}$ are the $(n,n)$th entries of $\bLambda_k$ and $\bGamma_k$, respectively. In summary, the DE method is detailed in \alref{alg:Deterministic}.

\begin{algorithm}[h]
\caption{DE Method}
\label{alg:Deterministic}
\begin{algorithmic}[1]
\Require The RIS phase shift matrix $\bPhi$, the power allocation matrices $\bLambda$, and threshold ${\varepsilon}$.
\For{$k=1$ to $K$}
\State Initialize $\bpsi_k^{(0)}$ and set iteration index $u = 0$.
\Repeat
\For{$m=1$ to $M$}
\State Calculate $\gamma_{k,m}^{(u+1)}$ by \eqref{eq:gamma} with $\bpsi_k^{(u)}$.
\EndFor
\State Obtain $\bgamma^{(u+1)}_k = \left[ \gamma_{k,1}^{(u+1)},\ldots,\gamma_{k,N_{\rmR}}^{(u+1)} \right]^T$.
\For{$n=1$ to $N_k$}
\State Calculate $\psi_{k,n}^{(u+1)}$ by \eqref{eq:psi} with $\bgamma^{(u+1)}_k$.
\EndFor
\State Obtain $\bpsi_k^{(u+1)} = \left[ \psi_{k,1}^{(u+1)},\ldots,\psi_{k,N_k}^{(u+1)} \right]^T$.
\State Set $u=u+1$.
\Until{$\left\|  {\bpsi}_k^{(u)} -  {\bpsi}_k^{(u-1)} \right\|\le \varepsilon$}
\State Use ${\bgamma}_k^{(u)}$ and ${\bpsi}_k^{(u)}$ to calculate $\bGamma_k$ and $\bPsi_k$ in \eqref{eq:T} and \eqref{eq:F}, respectively.
\EndFor
\State Set ${\bgamma}_k = {\bgamma}_k^{(u)}$ and $\bpsi_k = \bpsi_k^{(u)}$, $\forall k$, and use them to calculate $\overline{R} \left( \bLambda \right)$ in \eqref{eq:DE}.
\Ensure The DE-based system SE $\overline{R} \left( \bLambda \right)$ and the DE auxiliary variables $\bpsi_k$, $\forall k$.
\end{algorithmic}
\end{algorithm}

With the help of the derived DE expression in \eqref{eq:DE}, we then reformulate problem $\cP_{\bLambda}$ as
\begin{align}\label{eq:problem_Lambda}
\overline{\cP}_{\bLambda}:\quad \underset{\bLambda} \max \quad & \frac{ \overline{R} \left( \bLambda \right)   }{ \sum\limits_{k=1}^K \left( \xi_k \tr{\bLambda_k} + P_{\mathrm{c},k} \right) + P_{\mathrm{BS}} + N_{\rmR} P_{\mathrm{s}} } \ntb
{\mathrm{s.t.}}\quad
& \tr { \bLambda_k }  \le P_{\max,k}, \quad \lambdak \succeq \bzero,\quad \lambdak\; \mathrm{diagonal},\quad \forall k\in {\cal K}.
\end{align}
It is worth remarking that the DE-based system SE, $\overline{R} \left( \bLambda \right)$, is concave in terms of $\bLambda_k$, $\forall k$ \cite{Wen11On}.

\subsection{Transmit Power Allocation at UTs}
Problem $\overline{\cP}_{\bLambda}$ is intrinsically a classical fractional programming problem. Inspecting  $\overline{\cP}_{\bLambda}$, we find that its objective function exhibits a concave-convex ratio structure, where the numerator is concave and the denominator is convex. For the case of single-ratio concave-convex fractional problems, classical fractional programming techniques such as Charnes-Cooper algorithm and Dinkelbach's approach can be applied to obtain the optimal solution \cite{zappone2015energy}.

In this paper, the solution algorithm for $\overline{\cP}_{\bLambda}$ is developed by Dinkelbach's method, which is a kind of parametric algorithms. In particular, we introduce an auxiliary variable ${\eta}^{(\ell)}$, by which a sequence of easy-to-tackle subproblems is constructed. Specifically, the subproblem at the $\ell$th iteration of Dinkelbach's algorithm is given by
\begin{align}\label{eq:problem6}
\overline{\cP}^{(\ell)}_{\bLambda}:\quad \underset{\bLambda} \max \quad &  \overline{R} \left( \bLambda \right) - {\eta}^{(\ell)} \left( \sum\limits_{k=1}^K \left( \xi_k \tr{\bLambda_k} + P_{\mathrm{c},k} \right) + P_{\mathrm{BS}} + N_{\rmR} P_{\mathrm{s}} \right) \ntb
{\mathrm{s.t.}}\quad
& \tr { \bLambda_k }  \le P_{\max,k}, \quad \lambdak \succeq \bzero,\quad \lambdak\; \mathrm{diagonal},\quad \forall k\in {\cal K}.
\end{align}
We assume that the optimal solution of $\overline{\cP}^{(\ell-1)}_{\bLambda}$ is denoted as $\bLambda^{(\ell)} = \left\{\bLambda_k^{(\ell)}\right\}_{k=1}^K$, by which ${\eta}^{(\ell)}$ is iteratively updated as
\begin{align}\label{eq:update_eta}
{\eta}^{(\ell)} = \frac{ \overline{R}\left( \bLambda^{(\ell)} \right)}{ \sum\limits_{k=1}^K \left( \xi_k \tr{\bLambda^{(\ell)}_k} + P_{\mathrm{c},k} \right) + P_{\mathrm{BS}} + N_{\rmR} P_{\mathrm{s}}}.
\end{align}
After this transformation, the surrogate subproblem $\overline{\cP}^{(\ell)}_{\bLambda}$ is a standard concave program and we can obtain its optimal solution via classical convex optimization techniques \cite{Boyd04Convex}. In addition, the resultant sequence $\left\{ \bLambda^{(\ell)} \right\}_{\ell=0}^{\infty}$ will converge to the global optimum of $\overline{\cP}_{\bLambda}$ with a super-linear rate of convergence \cite{zappone2015energy}. More details about this approach is summarized in \alref{alg:Dinkelbach}.

\begin{algorithm}[h]
\caption{Dinkelbach's Algorithm}
\label{alg:Dinkelbach}
\begin{algorithmic}[1]
\Require The RIS phase shift matrix $\bPhi$ and threshold ${\varepsilon}$.
\State Initialize $\eta^{(0)}$ and set iteration index $\ell = 0$.
\Repeat
\State Solve the concave program $\overline{\cP}^{(\ell)}_{\bLambda}$ in \eqref{eq:problem6} and set $\bLambda^{(\ell+1)}$ as the intermediate solution.
\State Set $\ell=\ell+1$.
\State Update ${\eta}^{(\ell)}$ by \eqref{eq:update_eta}.
\Until{$\left|  {\eta}^{(\ell)} -  {\eta}^{(\ell-1)} \right|\le \varepsilon$}
\Ensure The optimal power allocation matrices $\bLambda^{(\ell)}$.
\end{algorithmic}
\end{algorithm}

\section{Optimization of RIS Phase Shift Matrix}\label{sec:Optimization_Phi}
In this section, we focus on the problem where $\bQ$ is fixed while the RIS phase shift matrix, $\bPhi$, needs to be optimized, which is characterized as
\begin{align}\label{eq:problem_Phi}
\underset{\bPhi} \max \quad & \frac{ \sum\limits_{k=1}^K \log_2 \det \left( \bI_{N_k} + \bGamma_k \bLambda_k \right) + \log_2 \det \left( \bI_M + \bPsi \right) - \sum\limits_{k=1}^K {\bgamma_k^T \bOmega_k \bpsi_k}   }{ \sum\limits_{k=1}^K \left( \xi_k \tr{\bLambda_k} + P_{\mathrm{c},k} \right) + P_{\mathrm{BS}} + N_{\rmR} P_{\mathrm{s}} } \ntb
{\mathrm{s.t.}}\quad
& \left| \phi_n \right| = 1,\quad n = 1,\ldots,N_{\mathrm{R}}.
\end{align}
Since the transmit covariance matrices, $\bQ_k$, of all UT $k$, are fixed, i.e., both the signal directions, $\bV_{k}$, $\forall k$, and the power allocation matrices, $\bLambda_k$, $\forall k$, are fixed, the denominator of the objective function in \eqref{eq:problem_Phi} is reduced to a constant. In addition, we update $\bPhi$ and $(\bgamma,\bpsi)$ in an iterative manner as shown in \cite{Wen11On}, i.e., fix the parameters $(\bgamma,\bpsi)$ when optimizing $\bPhi$ and then update $(\bgamma,\bpsi)$ by \eqref{eq:gamma} and \eqref{eq:psi}. Hence, only the second term of the DE expression in \eqref{eq:DE} is related to $\bPhi$, while the others are considered as constants with respect to $\bPhi$. Based on these observations, problem \eqref{eq:problem_Phi} is simplified into
\begin{align}\label{eq:problem_Phi_simplified}
\overline{\cP}_{\bPhi}:\quad \underset{\bPhi} \max \quad & C \left( \bPhi \right) = \log_2 \det \left( \bI_M + \sum\limits_{k=1}^K{\s \bH_1 \bPhi \bU_{2,k} \diag{ \bOmega_k \bpsi_k } \bU_{2,k}^H \bPhi^H \bH_1^H} \right)\ntb
{\mathrm{s.t.}}\quad
& \left| \phi_n \right| = 1,\quad n = 1,\ldots,N_{\mathrm{R}}.
\end{align}
It is not straightforward to solve $\overline{\cP}_{\bPhi}$ and the major challenges in this issue arise from the non-convexity of the objective function as well as the unit-modulus constraints. To facilitate the design of a computationally efficient algorithm, in the following, we first convert \eqref{eq:problem_Phi_simplified} into an equivalent MSE minimization problem and then provide an algorithm combining the BCD method with sequential convex optimization approaches.

\subsection{Equivalent MSE Minimization}
To proceed, we define $\bA=\sum\limits_{k=1}^K{ \bU_{2,k} \diag{ \bOmega_k \bpsi_k } \bU_{2,k}^H } \succeq \bzero$ for notational brevity. Then, the objective function in $\overline{\cP}_{\bPhi}$ is recast as
\begin{align}\label{eq:objective_A}
C \left( \bPhi \right) = \log_2 \det \left( \bI_M + \s \bH_1 \bPhi \bA \bPhi^H \bH_1^H \right).
\end{align}
To facilitate the understanding of the subsequent optimization, we treat $C\left( \bPhi \right)$ as the data rate of a hypothetical communication system where the received signal is modeled as
\begin{align}
\by_{\mathrm{c}} = \bH_1 \bPhi \bA^{1/2} \bx_{\mathrm{c}} + \bn_{\mathrm{c}},
\end{align}
with $ \bH_1 \bPhi \bA^{1/2}$ being the equivalent channel matrix. In addition, $\bx_{\mathrm{c}} \sim \mathcal{CN}(\bzero,\bI_{N_{\mathrm{R}}})$ and $\bn_{\mathrm{c}} \sim \mathcal{CN}(\bzero, \sigma^2 \bI_M)$ are the system input and the thermal noise, respectively. In this hypothetical system, the estimated signal adopting a linear decoder is given by
\begin{align}\label{eq:estimated_signal}
\widehat{\bx}_{\mathrm{c}} =  \bU_{\mathrm{c}}^H \by_{\mathrm{c}},
\end{align}
where $\bU_{\mathrm{c}} \in \C^{M \times N_{\mathrm{R}}}$ represents the receiving matrix. Assuming that $\bx_{\mathrm{c}}$ and $\bn_{\mathrm{c}}$ are independent, the MSE matrix is then computed as
\begin{align}\label{eq:MSE}
\bE_{\mathrm{c}} & \triangleq  \expectxn{ \left( \widehat{\bx}_{\mathrm{c}} - \bx_{\mathrm{c}} \right)\left( \widehat{\bx}_{\mathrm{c}} - \bx_{\mathrm{c}} \right)^H } \ntb
& = \expectxn{ \left[ \left( \bU_{\mathrm{c}}^H \bH_1 \bPhi \bA^{1/2} - \bI_{N_{\mathrm{R}}} \right) \bx_{\mathrm{c}} + \bU_{\mathrm{c}}^H \bn_{\mathrm{c}} \right] \left[ \left( \bU_{\mathrm{c}}^H \bH_1 \bPhi \bA^{1/2} - \bI_{N_{\mathrm{R}}} \right) \bx_{\mathrm{c}} + \bU_{\mathrm{c}}^H \bn_{\mathrm{c}} \right]^H } \ntb
& = \left( \bU_{\mathrm{c}}^H \bH_1 \bPhi \bA^{1/2} - \bI_{N_{\mathrm{R}}} \right)\left( \bU_{\mathrm{c}}^H \bH_1 \bPhi \bA^{1/2} - \bI_{N_{\mathrm{R}}} \right)^H + \sigma^2 \bU_{\mathrm{c}}^H \bU_{\mathrm{c}}.
\end{align}
Then, utilizing a result similar as that in \cite[Theorem 1]{shi2011an}, we introduce an auxiliary optimization matrix variable $\bW_{\mathrm{c}}\in \C^{N_{\mathrm{R}} \times N_{\mathrm{R}}}$ and apply it to establish a matrix-weighted MSE minimization problem as \cite{shi2011an}
\begin{align}\label{eq:MSE_minimization}
\overline{\cP}_{\mathrm{MSE}}:\quad \underset{\bW_{\mathrm{c}},\bU_{\mathrm{c}},\bPhi} \min \quad & h \left( \bW_{\mathrm{c}},\bU_{\mathrm{c}},\bPhi \right) \triangleq \tr{\bW_{\mathrm{c}} \bE_{\mathrm{c}}} - \log_2 \det \left( \bW_{\mathrm{c}} \right) \ntb
{\mathrm{s.t.}}\quad
& \left| \phi_n \right| = 1,\quad n = 1,\ldots,N_{\mathrm{R}},
\end{align}
which is equivalent to the rate maximization problem in \eqref{eq:problem_Phi_simplified}. Note that problem $\overline{\cP}_{\mathrm{MSE}}$ is easier to handle than the original problem $\overline{\cP}_{\bPhi}$, since the objective function in $\overline{\cP}_{\mathrm{MSE}}$ is convex in terms of each variable matrix ($\bW_{\mathrm{c}}$, $\bU_{\mathrm{c}}$ or $\bPhi$) when the other two variables are fixed. This structure enables the design of computationally efficient algorithms.

\subsection{BCD Method}
In practice, the number of optimization variables in $\overline{\cP}_{\mathrm{MSE}}$ can be large even in moderate system size, which remains a challenge in applying RISs in multiuser MIMO uplink systems. The BCD method is one of the fundamental techniques for handling large-size optimization problems. As a generalization of AO, the BCD method performs a similar procedure, i.e., alternatingly optimize one variable while regarding the others as constants. In the sequel, we propose a BCD-based method to handle the MSE minimization problem $\overline{\cP}_{\mathrm{MSE}}$ in \eqref{eq:MSE_minimization}. To be more specific, we minimize the objective function, $h \left( \bW_{\mathrm{c}},\bU_{\mathrm{c}},\bPhi \right)$, by sequentially updating $\bW_{\mathrm{c}}$, $\bU_{\mathrm{c}}$, and $\bPhi$.

The optimization of $\bW_{\mathrm{c}}$ with $\bU_{\mathrm{c}}$ and $\bPhi$ being fixed is straightforward. Since the minimization problem is convex over $\bW_{\mathrm{c}}$, the optimal solution can be derived by applying the first-order optimality condition of the Lagrangian function of $h \left( \bW_{\mathrm{c}},\bU_{\mathrm{c}},\bPhi \right)$ with respect to $\bW_{\mathrm{c}}$. In particular, the optimal $\bW_{\mathrm{c}}$ can be obtained in a closed-form as
\begin{align}\label{eq:optimal_W}
\bW_{\mathrm{c}}^{\mathrm{opt}} = \bE_{\mathrm{c}}^{-1}.
\end{align}
Similarly, for fixed $\bW_{\mathrm{c}}$ and $\bPhi$, the optimal $\bU_{\mathrm{c}}$ is given by
\begin{align}\label{eq:optimal_U}
\bU_{\mathrm{c}}^{\mathrm{opt}} = \left( \sigma^2 \bI_M + \bH_1 \bPhi \bA \bPhi^H \bH_1^H \right)^{-1}\bH_1 \bPhi \bA^{1/2}.
\end{align}
The results of $\bW_{\mathrm{c}}$ and $\bU_{\mathrm{c}}$ are clear and explicit. With given $\bW_{\mathrm{c}}$ and $\bU_{\mathrm{c}}$, the problem in \eqref{eq:MSE_minimization} is then reduced to
\begin{align}\label{eq:MSE_minimization_Phi}
\overline{\cP}_{\mathrm{MSE},\bPhi}:\quad \underset{\bPhi} \min \quad
& \tr{ \bPhi^H \bB \bPhi \bA  } - \tr{\bPhi^H \bC^H } - \tr{\bPhi \bC } \ntb
{\mathrm{s.t.}}\quad
& \left| \phi_n \right| = 1,\quad n = 1,\ldots,N_{\mathrm{R}},
\end{align}
where $\bB = \bH_1^H \bU_{\mathrm{c}} \bW_{\mathrm{c}} \bU_{\mathrm{c}}^H \bH_1 \in \C^{N_{\mathrm{R}} \times N_{\mathrm{R}}}$ and $\bC = \bA^{1/2}\bW_{\mathrm{c}}\bU_{\mathrm{c}}^H \bH_1 \in \C^{N_{\mathrm{R}} \times N_{\mathrm{R}}} $. Notice that $\bW_{\mathrm{c}}$ and $\bU_{\mathrm{c}}$ are deterministic during the process of optimizing $\bPhi$. In particular, $\log_2 \det \left( \bW_{\mathrm{c}} \right)$ and $\tr{ \sigma^2 \bW_{\mathrm{c}} \bU_{\mathrm{c}}^H \bU_{\mathrm{c}} }$ are regarded as constant terms and therefore can be omitted, which results in the reduced minimization problem in \eqref{eq:MSE_minimization_Phi}.

Recall that $\bPhi = \diag{\phi_1,\ldots,\phi_{N_{\rmR}}}$, where $\abs{\phi_n}=1,\forall n$. For the convenience of the subsequent expressions, we define $\bphi = \left[\phi_1,\ldots,\phi_{N_{\rmR}}\right]^T$ and $\bc = \left[ \left[\bC\right]_{1,1}, \ldots, \left[\bC\right]_{N_{\mathrm{R}},N_{\mathrm{R}}} \right]^T$ as the vectors collecting the diagonal components of $\bPhi$ and $\bC$, respectively. Equipped with these notations, we have
\begin{subequations}
\begin{align}\label{eq:first}
& \tr{ \bPhi^H \bB \bPhi \bA  } = \bphi^H \left( \bB \odot \bA^T \right)  \bphi,\\ \label{eq:second_third}
& \tr{\bPhi^H \bC^H }  = \bc^H \bphi^*, \quad \tr{\bPhi \bC } = \bphi^T\bc,
\end{align}
\end{subequations}
where the equation in \eqref{eq:first} follows from the matrix identity in \cite[Eq. (1.10.6)]{Zhang2017Matrix}. Accordingly, problem \eqref{eq:MSE_minimization_Phi} can be equivalently expressed as
\begin{align}\label{eq:minimization_Phi}
\underset{\bphi} \min \quad & g\left(\bphi\right) = \bphi^H \left( \bB \odot \bA^T \right) \bphi - 2 \Real{\bphi^H\bc^*}  \ntb
{\mathrm{s.t.}}\quad
& \left| \phi_n \right| = 1,\quad n = 1,\ldots,N_{\mathrm{R}}.
\end{align}

\subsection{MM Technique}

The optimization of $\bphi$ in problem \eqref{eq:minimization_Phi} is challenging since the unit-modulus constraints exhibit non-convexity. In the following, we resort to the MM technique, which belongs to the sequential convex optimization approaches, in order to obtain a suboptimal solution. Developing tractable surrogate subproblems is the key and decides the effectiveness of the MM technique. Adhere to this idea, we aim to approximate the objective by its surrogate function such that the constraint in \eqref{eq:minimization_Phi} can be handled. To this end, we start from the following lemma.
\begin{lemma} \label{lemma:upper}
Suppose $\bS$ and $\bL$ are both Hermitian matrices and $\bL \succeq \bS$. For an arbitrarily given $\bphi^{(0)}$, we have
\begin{align} \label{eq:upper}
\bphi^H \bS \bphi \le \bphi^H \bL \bphi - 2 \Real{ \bphi^H \left( \bL - \bS \right)\bphi^{(0)} } + \left(\bphi^{(0)}\right)^H \left( \bL - \bS \right) \bphi^{(0)}.
\end{align}
\end{lemma}

\begin{IEEEproof}
Since $\bL \succeq \bS$, the constructed $\left( \bL - \bS \right)$ is essentially positive semi-definite. Elaborating the fact that $\left\| \left( \bL - \bS \right)^{1/2}\bphi - \left( \bL - \bS \right)^{1/2}\bphi^{(0)} \right\|^2  \ge 0$, we have
\begin{align} \label{eq:prove_lemma}
\bphi^H \left( \bL - \bS \right) \bphi + \left(\bphi^{(0)}\right)^H \left( \bL - \bS \right) \bphi^{(0)} - 2 \Real{\bphi^H \left( \bL - \bS \right) \bphi^{(0)}} \ge 0.
\end{align}
Then, by means of insulating the term $\bphi^H \bS \bphi$ in \eqref{eq:prove_lemma}, we obtain the inequality in \eqref{eq:upper}. This concludes the proof.
\end{IEEEproof}

Inspired by the inequality in \emph{\lmref{lemma:upper}}, we denote $\bS = \bB \odot \bA^T \in \C^{N_{\mathrm{R}} \times N_{\mathrm{R}}} $, which can be verified to be positive semi-definite according to \cite{Zhang2017Matrix} since $\bB$ and $\bA$ are both positive semi-definite. Also, we denote $\bL = \lambda_{\max} \bI_{N_{\mathrm{R}}} $ with $\lambda_{\max}$ being the maximum eigenvalue of $\bS$, so that $\bL - \bS$ is positive semi-definite. Applying \emph{\lmref{lemma:upper}}, we then establish an upper bound of the objective function in \eqref{eq:minimization_Phi} as
\begin{align} \label{eq:g_upperbound}
g\left(\bphi\right) & \le \widetilde{g} \left(\bphi | \bphi^{(i)}\right) \ntb
& = \bphi^H \bL \bphi - 2 \Real{ \bphi^H \left( \bL - \bS \right)\bphi^{(i)} } + \left(\bphi^{(i)}\right)^H \left( \bL - \bS \right) \bphi^{(i)} - 2 \Real{\bphi^H\bc^*},
\end{align}
where $i$ is the iterative index and $\bphi^{(i)}$ denotes the minimizer at the $(i-1)$th iteration of the MM procedure. Then, with the aid of the reconstructed objective function, we obtain the following surrogate subproblems as
\begin{subequations}\label{eq:minimization_Phi_MM}
\begin{align} \label{eq:objective}
\overline{\cP}_{\mathrm{MSE},\bPhi}^{(i)}:\quad \underset{\bphi} \min \quad & \widetilde{g} \left(\bphi | \bphi^{(i)}\right)  \\ \label{eq:constraint}
{\mathrm{s.t.}}\quad
& \left| \phi_n \right| = 1,\quad n = 1,\ldots,N_{\mathrm{R}}.
\end{align}
\end{subequations}
Since the moduli of $\phi_n$, $\forall n$, are constrained to be unity, we have $\bphi^H \bphi = \bI_{N_{\mathrm{R}}} $ and $\bphi^H \bL \bphi = \lambda_{\max} N_{\mathrm{R}}$. Moreover, omitting the constant terms irrespective of $\bphi$, we arrive at an equivalence of problem \eqref{eq:minimization_Phi_MM} which is given by
\begin{align} \label{eq:phi_final}
\underset{\bphi} \max \quad & \Real{ \bphi^H \balpha^{(i)} } \ntb
{\mathrm{s.t.}}\quad
& \left| \phi_n \right| = 1,\quad n = 1,\ldots,N_{\mathrm{R}},
\end{align}
where $\balpha^{(i)} = \left( \lambda_{\max} \bI_{N_{\mathrm{R}}} - \bS\right)\bphi^{(i)} + \bc^*$. We denote the $n$th element of vector $\balpha^{(i)}$ by $\alpha^{(i)}_n = \left|\alpha^{(i)}_n\right| \mathrm{e}^{\jmath \beta^{(i)}_n}$. Then, it is trivial to obtain the optimal solution to problem \eqref{eq:phi_final}, which is
\begin{align} \label{eq:phi_solution}
\phi_n^{(i+1)} = \mathrm{e}^{\jmath \beta^{(i)}_n},\quad n = 1,\ldots,N_{\mathrm{R}}.
\end{align}
It is revealed from \eqref{eq:phi_solution} that to minimize the objective function in \eqref{eq:minimization_Phi_MM}, the phase of each $\phi_n$ should be aligned with that of the corresponding element of $\balpha^{(i)}$. Denoting the minimum objective value of $\overline{\cP}_{\mathrm{MSE},\bPhi}^{(i)}$ as $\widetilde{g}^{(i)}_{\min}$ and $\bPhi^{(i)} = \diag{\bphi^{(i)}}$, we can obtain the following results.
\begin{prop}\label{prop:mm_convergence}
The minimum objective value sequence, $\left\{ \widetilde{g}^{(i)}_{\min} \right\}_{i=0}^{ \infty }$, output by $\overline{\cP}_{\mathrm{MSE},\bPhi}^{(i)}$ is monotonically non-increasing and convergent. Additionally, the sequence of the corresponding RIS phase shift matrices, $\left\{ \bPhi^{(i)} \right\}_{i=0}^{ \infty }$, also converges, with each limit point being a local minimizer of the original problem $\overline{\cP}_{\mathrm{MSE},\bPhi}$. Lastly, the resulting point of $\left\{ \bPhi^{(i)} \right\}_{i=0}^{ \infty }$ fulfills the first-order optimality conditions of problem $\overline{\cP}_{\mathrm{MSE},\bPhi}$.
\end{prop}
\begin{IEEEproof}
See \appref{app:B}.
\end{IEEEproof}

To conclude this section, the MM-based approach proposed to handle problem \eqref{eq:minimization_Phi} is summarized in \alref{alg:MM}.

\begin{algorithm}[h]
\caption{MM-based Phase Shift Optimization Method}
\label{alg:MM}
\begin{algorithmic}[1]
\Require Feasible $\bphi^{(0)}$, iterative index $i=0$, and threshold ${\varepsilon}$.
\State Calculate the objective value $g\left(\bphi^{(i)}\right)$ in \eqref{eq:minimization_Phi}.
\Repeat
\State Calculate $\balpha^{(i)} = \left( \lambda_{\max} \bI_{N_{\mathrm{R}}} - \bB \odot \bA^T \right)\bphi^{(i)} + \bc^*$.
\State Solve problem $\overline{\cP}_{\mathrm{MSE},\bPhi}^{(i)}$ in \eqref{eq:minimization_Phi_MM} with its optimal solution $\bphi^{(i+1)}$ given by \eqref{eq:phi_solution}.
\State Calculate $g\left(\bphi^{(i+1)}\right)$.
\State Set $i = i + 1$.
\Until{$\left| g\left(\bphi^{(i)}\right) -  g\left(\bphi^{(i-1)}\right) \right|\le \varepsilon$}
\Ensure The RIS phase shift vector $\bphi^{(i)}$.
\end{algorithmic}
\end{algorithm}

\section{Overall Algorithm}\label{sec:overallopt}

\subsection{GEE Maximization}
The above Sections \ref{sec:Optimization_Q} and \ref{sec:Optimization_Phi} provide the solutions for $\bQ$ and $\bPhi$, respectively. Combining all the adopted approaches together forms the overall solution methodology for the GEE maximization problem. In particular, we present the detailed description of the GEE maximization algorithm for the considered RIS-assisted multiuser MIMO uplink transmission in \alref{alg:GEE}, where the BCD method in step 8 is detailed in \alref{alg:BCD}.

\begin{algorithm}[h]
\caption{AO-based GEE Maximization Algorithm}
\label{alg:GEE}
\begin{algorithmic}[1]
\Require Feasible $\bLambda^{(0)}$, $\bPhi^{(0)}$, iterative index $t=0$, and threshold ${\varepsilon}$.
\Repeat
\State Calculate the DE expression $\overline{R} \left( \bLambda^{(t)} \right)$ by \alref{alg:Deterministic}.
\State Update $\bQ$ with given $\bPhi^{(t)}$:
\State \ \ Solve $\overline{\cP}_{\bLambda}$ by \alref{alg:Dinkelbach} and set the optimal solution as $\bLambda_k^{(t+1)}, \ k = 1,\ldots,K$.
\State \ \ Obtain $\bQ_k^{(t+1)} = \bV_{2,k}^H  \bLambda_k^{(t+1)} \bV_{2,k}, \ k = 1,\ldots,K$.
\State Update $\bPhi$ with given $\bQ^{(t+1)}$:
\State \ \ Calculate the DE auxiliary variables $\bpsi_k$, $\forall k$, using $\bPhi^{(t)}$ and $\bLambda^{(t+1)}$ by \alref{alg:Deterministic}.
\State \ \ Solve problem $\overline{\cP}_{\bPhi}$ in \eqref{eq:problem_Phi_simplified} via solving the equivalent problem $\overline{\cP}_{\mathrm{MSE}}$ using the BCD-based \alref{alg:BCD} and set the intermediate solution as $\bPhi^{(t+1)}$.
\State Set $t = t + 1$.
\Until{$\left|  \mathrm{GEE}\left( \bQ^{(t)},\bPhi^{(t)} \right) -  \mathrm{GEE}\left( \bQ^{(t-1)},\bPhi^{(t-1)} \right) \right|\le \varepsilon$}
\Ensure Transmit covariance matrices $\bQ^{(t)}$ and the RIS phase shift matrix $\bPhi^{(t)}$.
\end{algorithmic}
\end{algorithm}

\begin{algorithm}[h]
\caption{BCD Method}
\label{alg:BCD}
\begin{algorithmic}[1]
\Require Feasible $\bW_{\mathrm{c}}^{(0)}$, $\bU_{\mathrm{c}}^{(0)}$, $\bPhi^{(0)}$, iterative index $s=0$, and threshold ${\varepsilon}$.
\State Calculate $h \left( \bW_{\mathrm{c}}^{(s)},\bU^{(s)}_{\mathrm{c}},\bPhi^{(s)} \right)$ in \eqref{eq:MSE_minimization}.
\Repeat
\State Update $\bW_{\mathrm{c}}$ with given $\bU_{\mathrm{c}}^{(s)}$ and $\bPhi^{(s)}$:
\State \ \ Calculate $\bE_{\mathrm{c}}$ in \eqref{eq:MSE} with $\bU_{\mathrm{c}}^{(s)}$ and $\bPhi^{(s)}$.
\State \ \ Obtain $\bW_{\mathrm{c}}^{(s+1)} = \bE_{\mathrm{c}}^{-1}$.
\State Update $\bU_{\mathrm{c}}$ with given $\bW_{\mathrm{c}}^{(s+1)}$ and $\bPhi^{(s)}$:
\State \ \ Obtain $\bU_{\mathrm{c}}^{(s+1)} = \left( \sigma^2 \bI_M + \bH_1 \bPhi^{(s)} \bA (\bPhi^{(s)})^H \bH_1^H \right)^{-1}\bH_1 \bPhi^{(s)} \bA^{1/2}$.
\State Update $\bPhi$ with given $\bW_{\mathrm{c}}^{(s+1)}$ and $\bU_{\mathrm{c}}^{(s+1)}$:
\State \ \ Solve the problem in \eqref{eq:minimization_Phi} using the MM-based \alref{alg:MM} and set the optimal solution as $\bphi^{(s+1)}$.
\State \ \ Obtain $\bPhi^{(s+1)} = \diag{\bphi^{(s+1)}}$.
\State Calculate $h \left( \bW_{\mathrm{c}}^{(s+1)},\bU^{(s+1)}_{\mathrm{c}},\bPhi^{(s+1)} \right)$ in \eqref{eq:MSE_minimization}.
\State Set $s = s + 1$.
\Until{$\left|  h \left( \bW_{\mathrm{c}}^{(s)},\bU^{(s)}_{\mathrm{c}},\bPhi^{(s)} \right) -  h \left( \bW_{\mathrm{c}}^{(s-1)},\bU_{\mathrm{c}}^{(s-1)},\bPhi^{(s-1)} \right) \right|\le \varepsilon$}
\Ensure The RIS phase shift matrix $\bPhi^{(s)}$.
\end{algorithmic}
\end{algorithm}

\subsection{SE Maximization}\label{sec:SEMaximization}

In the previous discussions, we focused on the considered GEE maximization problem in \eqref{eq:problem_Q_phi}. Although the overall algorithm is designed for maximizing the system GEE, we can straightforwardly specialize the proposed approach to the case of maximizing the system SE. In fact, inspecting the objective function of problem \eqref{eq:problem_Q_phi}, we find that the system SE appears as just the numerator of the GEE. Therefore, if we consider the numerator only, the GEE maximization in \alref{alg:GEE} is reduced to handle the special case that maximize the system SE. To perform this modification, we just need to set $\xi_k = 0$ for all UTs, so that the denominator of the system GEE is degenerated into a constant. Note that this setting has a great impact on the optimization of the power allocation matrices, $\bLambda_k$, $\forall k$. Specifically, compared with the fractional, non-convex, and complicated GEE maximization, the SE maximization is a non-fractional, convex, and simple problem with respect to $\bLambda_k$, $\forall k$, and thus can be tackled after just one iteration in \alref{alg:Dinkelbach} without the use of Dinkelbach's method.

\subsection{Convergence and Complexity Analysis}

It is shown in \alref{alg:GEE} that $\bQ$ and $\bPhi$ are alternatingly optimized. For the solution approach obtaining $\bQ$, we iteratively optimize the eigenmatrix, $\bV$, and the power allocation matrix, $\bLambda$. As shown in \emph{\propref{theorem:beam_domain_optimal}}, $\bV$ has a closed-form optimal solution. Meanwhile, we optimize $\bLambda$ by Dinkelbach's method. Thus, the result converges to the global optimum of the fractional program $\overline{\cP}_{\bLambda}$ in \eqref{eq:problem_Lambda} \cite{zappone2015energy}. Consequently, $\bQ$ converges and would not decrease the system GEE value at each iteration in \alref{alg:GEE}.
In addition, the iterative MSE minimization approach conceived for $\bPhi$ in \alref{alg:BCD} is based on the BCD method and its convergence is guaranteed from \cite[Theorem 3]{shi2011an}.
Hence, the developed approach for optimizing $\bPhi$ also converges and will not decrease the system GEE value at each iteration in \alref{alg:GEE}. Based on the above facts, both solutions of $\bQ$ and $\bPhi$ will not decrease the objective value in $\cP_{\bQ,\bPhi}$, i.e., $\mathrm{GEE}\left( \bQ^{(t)},\bPhi^{(t)} \right) \ge  \mathrm{GEE}\left( \bQ^{(t-1)},\bPhi^{(t-1)} \right)$, where $t$ is the iteration index of AO. Hence, the convergence of the overall methodology for alternatingly optimizing $\bQ$ and $\bPhi$ in the AO-based GEE maximization in \alref{alg:GEE} is guaranteed.

After the convergence analysis of these algorithms, we turn our attention to discussing their computational complexity.
The main structure of the overall \alref{alg:GEE} is built upon AO, which requires a total of $I_{\mathrm{AO}}$ iterations. More specifically, due to the fast convergence rate of the DE method \cite{Couillet11Random} in \alref{alg:Deterministic}, the per-iteration complexity in \alref{alg:GEE} is mainly composed of the complexity of \alref{alg:Dinkelbach} for optimizing $\bLambda$ and the complexity of \alref{alg:BCD} for optimizing $\bPhi$.
For \alref{alg:Dinkelbach}, there is a total of $I_{\mathrm{D}}$ iterations included in the Dinkelbach's method and each iteration needs to tackle a convex program with $N$ variables, whose complexity is polynomial in terms of the number of variables \cite{ben2001lectures}. Hence, the complexity of \alref{alg:Dinkelbach} can be asymptotically estimated as $\cO(I_{\mathrm{D}} N^p)$ where the value of $I_{\mathrm{D}}$ is very small thanks to the super-linear convergence rate of Dinkelbach's method \cite{zappone2015energy} and $1 \le p \le 4$ for standard convex program solutions \cite{huang2019reconfigurable}. For \alref{alg:BCD}, we assume that the BCD method requires to perform $I_{\mathrm{BCD}}$ iterations, each comprises three major optimizations in terms of $\bW_{\mathrm{c}}$, $\bU_{\mathrm{c}}$, and $\bPhi$, respectively. It is clear to show that the complexity of computing the optimal results of $\bW_{\mathrm{c}}$ and $\bU_{\mathrm{c}}$, respectively given in \eqref{eq:optimal_W} and \eqref{eq:optimal_U}, is evaluated as $\cO(M^3)$ and $\cO(N_{\rmR}^3)$. Then, we analyze the complexity of the MM-based optimization of $\bPhi$ in \alref{alg:MM}. At the start of MM, it is necessary to obtain $\lambda_{\max}$, i.e., the maximum eigenvalue of the $N_{\rmR} \times N_{\rmR}$ matrix $\bS$, whose complexity is $\cO(N_{\rmR}^3)$. Suppose that the MM technique requires $I_{\mathrm{MM}}$ iterations to converge in total. The complexity of each iteration mainly depends on the computation of $\balpha^{(i)}$ in step 3 of \alref{alg:MM} and the corresponding complexity is given by $\cO(N_{\rmR}^2)$. Therefore, assuming that $N_{\rmR} > M$, the complexity of evaluating $\bPhi$ is approximated as $\cO(N_{\rmR}^3 + I_{\mathrm{MM}}N_{\rmR}^2)$. Hence, the complexity of obtaining the optimal $\bW_{\mathrm{c}}$ and $\bU_{\mathrm{c}}$ is negligible compared with that of optimizing $\bPhi$ in the BCD method, whose complexity is given by $\cO(I_{\mathrm{BCD}}(N_{\rmR}^3 + I_{\mathrm{MM}}N_{\rmR}^2))$. Putting together the above analyses, the overall complexity of the AO-based GEE maximization \alref{alg:GEE} is estimated as
$\cO( I_{\mathrm{AO}}(I_{\mathrm{D}} N^p + I_{\mathrm{BCD}}(N_{\rmR}^3 + I_{\mathrm{MM}}N_{\rmR}^2)))$, which is in polynomial time.

\section{Numerical Results}\label{sec:numerical_results}
In this section, we provide numerical results to appraise the performance of the proposed approach for our considered RIS-assisted multiuser MIMO uplink transmission. Throughout the simulations, the channel realizations are generated as follows. Regarding the large scale fading, we assume that all the composite UT-RIS-BS channels, i.e., $\bH_{1} \bH_{2,k}$, $\forall k$, exhibit the same path loss $-120$ dB for illustration \cite{Guo2019weighted}. Meanwhile, for the small scale fading of the UT-to-RIS and RIS-to-BS channels, we consider the suburban macro propagation environment where the primary statistical channel parameters are based on the 3GPP spatial channel model \cite{Salo05MATLAB}. Unless further specified, the simulation parameters are given as follows \cite{huang2019reconfigurable,Bjornson15Optimal}: number of UTs $K=8$, number of UT antennas $N_k = 4$, $\forall k$, number of BS antennas $M = 16$, number of reflecting units $N_{\rmR} = 32$, system bandwidth $W = 10$ MHz, background noise variance at the BS $\sigma^2 = -96$ dBm, amplifier efficiency factor $\rho_k = 0.3$, $\forall k$, i.e., $\xi_k = 1/0.3$, $\forall k$, static circuit power of each UT $P_{\mathrm{c},k} = 20$ dBm, $\forall k$, hardware dissipated power at the BS $P_{\mathrm{BS}} = 39$ dBm, static power per phase shifter at the RIS $P_{\mathrm{s}} = 10$ dBm, and maximum tolerance for algorithm convergence is ${\varepsilon} = 10^{-4}$. In addition, we assume equal individual maximum power constraints for all UTs, i.e., $P_{\max,k} = P_{\max}$, $\forall k$.

\subsection{Impact of Maximum Transmit Power}
\begin{figure}
\centering
\includegraphics[width=0.6\textwidth]{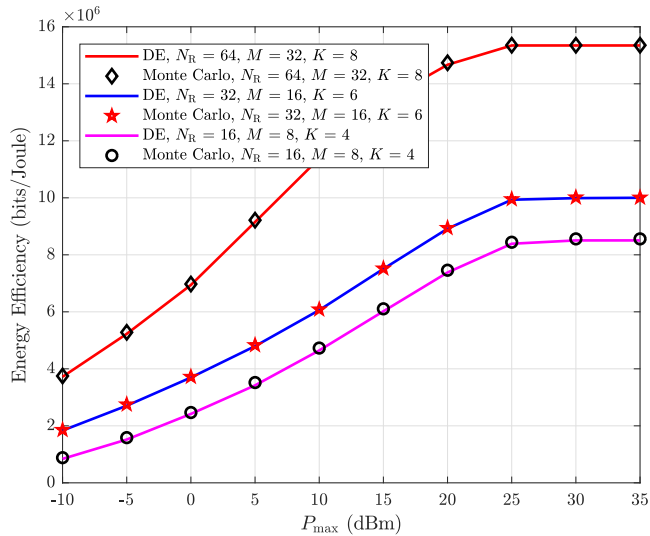}
\caption{Comparison between the DE and Monte-Carlo results of the GEE.}
\label{fig:ergodic_VS_DE}
\end{figure}

\figref{fig:ergodic_VS_DE} sketches the system GEE performance versus the maximum transmit power $P_{\max}$. We consider three cases with configurations given by: case 1) $N_{\rmR} = 16$, $M=8$, and $K=4$; case 2) $N_{\rmR} = 32$, $M=16$, and $K=6$; case 3) $N_{\rmR}=64$, $M=32$, and $K=8$. It can be seen that the system GEE achieved by the proposed approach first increases rapidly with increasing $P_{\max}$, and then becomes a constant when $P_{\max}$ is larger than a certain threshold value. This is a direct result of the fact that the system GEE is not a monotonically-increasing function with respect to $P_{\max}$. Instead, GEE is maximized by a finite but sufficient amount of transmit power. Once the maximum GEE is achieved, the proposed algorithm clips the transmit power even though there is still transmit power available which causes the saturation.

In addition, to verify the accuracy of the derived analytical DE expression, \figref{fig:ergodic_VS_DE} also compares it with the ergodic system GEE, which is evaluated through the computationally expensive Monte Carlo method. As can be seen from \figref{fig:ergodic_VS_DE}, the differences between the Monte Carlo results and the DE results are almost negligible in all the considered cases, even in that with moderate numbers of antennas. The simulation results illustrate that the proposed DE expressions are accurate to estimate the ergodic objective values. Thus, we confirm the effectiveness and validity of the proposed DE-based approach for resource allocation in the RIS-assisted multiuser MIMO uplink system with partial CSI.

\subsection{Comparison with the SE Maximization Approach}
\begin{figure*}[!t]
\centering
\subfloat[]{\centering\includegraphics[width=0.48\textwidth]{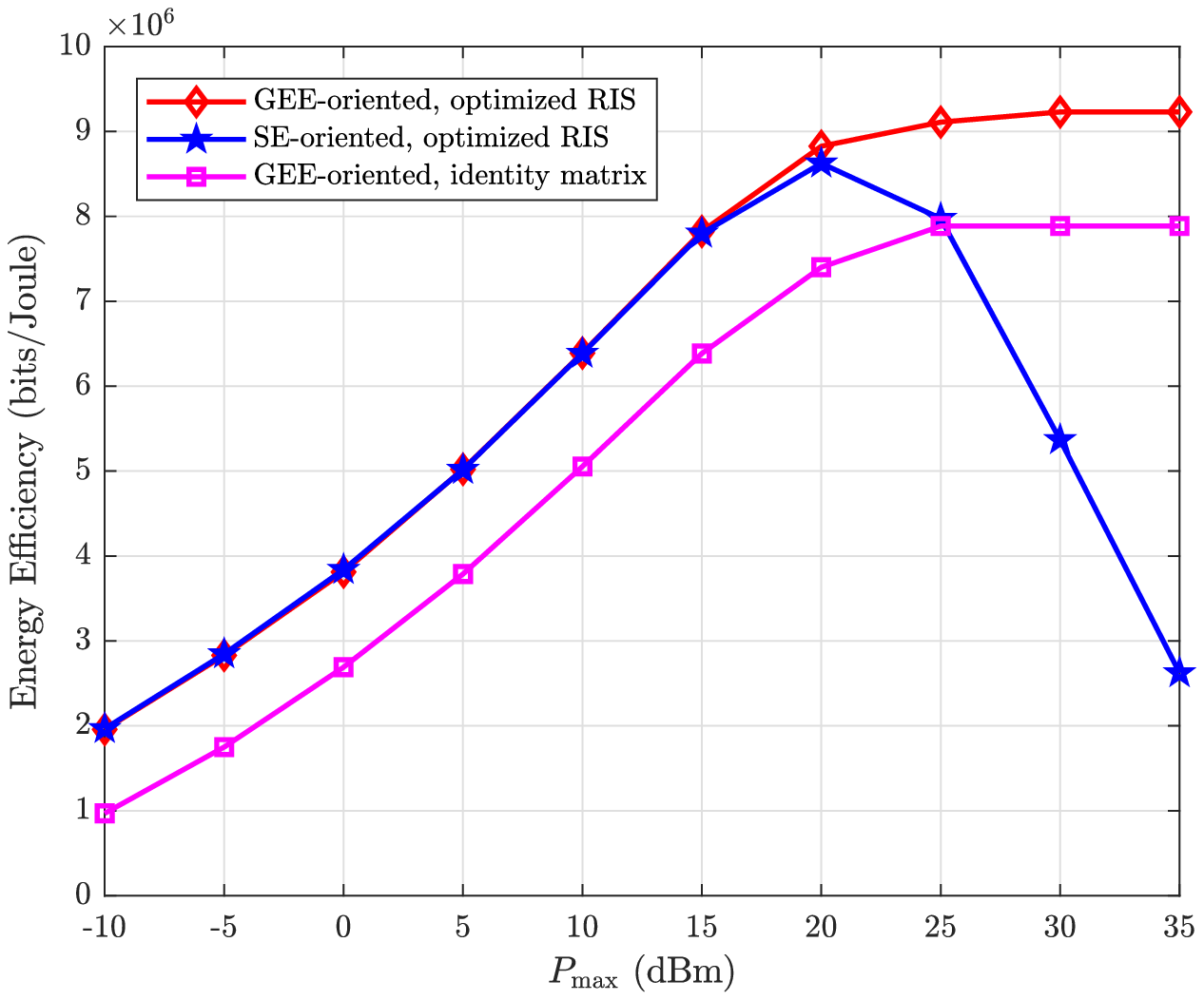}
\label{fig:EEvsSE_EE}}
\hfill
\subfloat[]{\centering\includegraphics[width=0.48\textwidth]{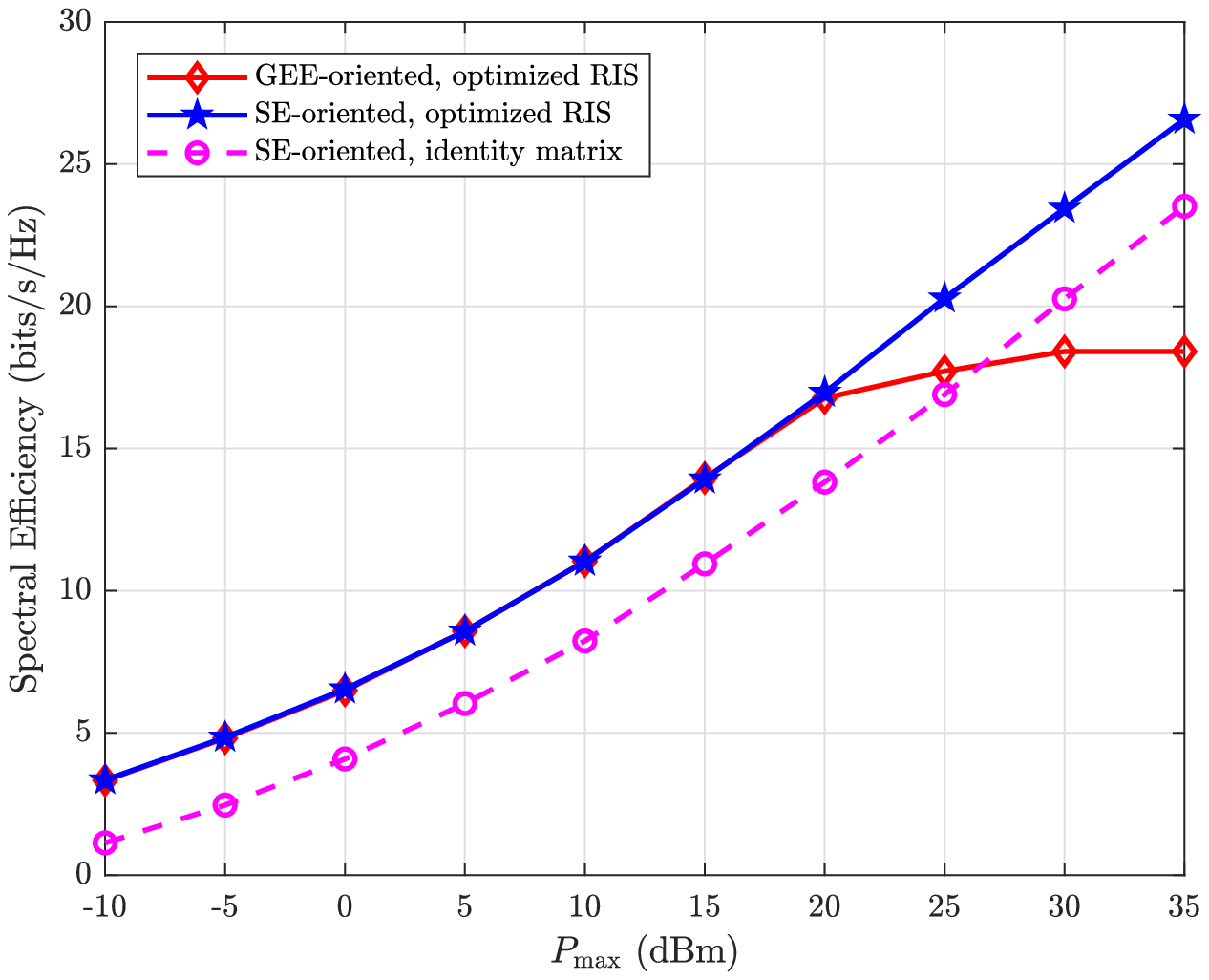}
\label{fig:EEvsSE_SE}}
\caption{Comparison of the GEE performance and SE performance versus $\Pmax$ with the aims of maximizing GEE and SE using \alref{alg:GEE}. (a) GEE performance; (b) SE performance.}
\label{fig:Comparison}
\end{figure*}

In \figref{fig:Comparison}, we compare the performances of the GEE-oriented approach with the SE-oriented one. The latter approach aims to maximize the system SE and is a special case of \alref{alg:GEE}, where $\xi_k$ for all $k$ is set to be zero as described in \secref{sec:SEMaximization}. The GEE and SE performances versus $P_{\max}$ are presented in Figs. 3(a) and 3(b), respectively. As depicted, when $P_{\max} \le 15$ dBm, these two approaches perform almost identically in terms of both GEE and SE. The results exhibit that at low transmit power levels, adopting GEE or SE as the system design criterion yields similar resource allocation. This is owing to the fact that in low $P_{\max}$ regime, both GEE and SE increase with $P_{\max}$ as the circuit power consumption dominates the objective function in \eqref{eq:problem_Q}. In other words, transmission exhausting the total power budget is energy-efficient and GEE maximization degenerates to SE maximization. However, the above two approaches perform substantially differently at high transmit power regimes, which is $P_{\max} \ge 15$ dBm in both subfigures. Observed from Fig. 3(b), the system SE achieved by the SE maximization approach keeps increasing with $P_{\max}$ while that achieved by the GEE-oriented one tends to be a constant. However, it is shown in Fig. 3(a) that the GEE-oriented approach remains as a constant while achieving a substantially higher GEE than that of the SE-oriented one. As mentioned before, there exists a saturation point of the optimal transmit power for maximizing GEE, thus any power exceeds the threshold is redundant and will only decrease the system GEE. In contrast, for the SE-oriented one, it always requires full power budget to maximize the SE, and thus an exceedingly larger transmit power is consumed, which decreases the GEE in high transmit power regimes.

To validate the benefits of utilizing RISs to enhance the system performances, we also compare the GEE performance of the RIS-assisted case to that with a fixed phase shift matrix, i.e., $\bPhi = \bI_{N_\rmR}$. In the latter case, only the optimization of the transmit covariance matrices, $\bQ_k$, of all UTs, is performed, which can be accomplished by means of \emph{\propref{theorem:beam_domain_optimal}}  together with the power allocation approach in \alref{alg:Dinkelbach}. As expected, the absence of the optimized RIS phase shift matrix $\bPhi$ leads to a degradation of the system GEE compared to the case where an optimized RIS is adopted, which is shown in Fig. 3(a). Moreover, we also plot the SE performance of the RIS-assisted system in comparison with the case of $\bPhi = \bI_{N_\rmR}$ in Fig. 3(b). It is intuitive and reasonable to see that the RIS-assisted one outperforms the other in terms of system SE. Generally, these results demonstrate the benefits of the RIS structure offering significant gains in both GEE and SE.

\subsection{Comparison with Other Schemes}

To further verify the GEE advantages brought by the deployment of RISs, we compare the RIS-assisted system with other schemes.
As the direct UT-to-BS channels are not available in the considered system, the performance of the conventional multiuser MIMO uplink systems without RIS cannot be guaranteed. Then, we consider a more relevant baseline scheme where the RIS is substituted by an amplify-and-forward (AF) relay equipped with $N_{\rmR}$ transmit and receive antennas, respectively. Note that AF is a widely adopted protocol as decoding is not required at the relay, which allows a more efficient implementation in practice \cite{Zappone2014energye}.
For the considered AF relay baseline case (assuming full-duplex operation with perfect self-interference cancellation), we model the operations at the AF relay by a $N_{\rmR} \times N_{\rmR}$ complex-valued matrix $\bF$, which is constrained by a maximum relay power budget. Note that this baseline scheme not only actively amplifies the desired signals, but also amplifies the receiver noise at the relay node, which does not happen in the RIS-assisted system. Consequently, the received signals at the relay and the BS can be expressed as
\begin{align}\label{eq:received_signal_AF}
\by_{\AF} & = \sum\limits_{k=1}^K { \bH_{2,k} \bx_k } + \bn_{\AF}, \\ \label{eq:received_signal_BS}
\by_{\BS} & = \bH_1 \bF \by_{\AF}  + \bn = \sum\limits_{k=1}^K {\bH_1 \bF \bH_{2,k} \bx_k } + \bH_1 \bF \bn_{\AF} + \bn,
\end{align}
respectively, where $\bn_{\AF} \sim \mathcal{CN}(\bzero,\sigma^2_{\AF} \bI_{N_{\rmR}})$ represents the thermal noise at the relay. Denote the aggregate interference-plus-noise as $\bn' = \bH_1 \bF \bn_{\AF} + \bn$, which is treated as Gaussian noise for a worst-case design \cite{Hassibi2003How}. Then, the ergodic SE of the relay-assisted multiuser MIMO system in the uplink is given by
\begin{align}\label{eq:ergodic_rate_AF}
R_{\AF} =  \expect { \log_2 \det  \left( \bK + \sum\limits_{k=1}^K \bH_1 \bF \bH_{2,k} \bQ_k \bH^H_{2,k}  \bF^H \bH^H_1 \right) } - \log_2 \det \left( \bK \right),
\end{align}
where $\bK$ is the covariance matrix of $\bn'$ and can be expressed as
\begin{align}\label{eq:K_AF}
\bK = \sigma^2_{\AF} \bH_1 \bF \bF^H \bH_1^H + \sigma^2 \bI_M \in \C ^{ M \times M }.
\end{align}
In addition, the energy consumption of the AF relay-assisted system is modeled as
\begin{align}\label{eq:power_consumption_AF}
P_{\AF,\tot} = \sum\limits_{k=1}^K \left( \xi_k \tr{\bQ_k} + P_{\mathrm{c},k} \right) + P_{\mathrm{BS}} + \xi_{\AF} P_{\AF}(\bF) +  N_{\rmR} P_{\mathrm{s},\AF},
\end{align}
which is similar to that of the RIS-assisted one, except for an additional transmit power $P_{\AF}(\bF)$ consumed by the relay for signal amplification. In addition, in \eqref{eq:power_consumption_AF}, $\xi_{\AF}$ is related to the relay power amplifier efficiency, $P_{\mathrm{s},\AF}$ denotes the power dissipated by each transmit antenna at the relay, and the relay total transmit power $P_{\AF}(\bF)$ is given by
\begin{align}\label{eq:transmit_power_AF}
P_{\AF}(\bF) & = \expect{\tr{ \bF \left( \sum\limits_{k=1}^{K} { \bH_{2,k} \bQ_k \bH^H_{2,k} } + \sigma^2_{\AF}\bI_{N_{\rmR}} \right) \bF^H }}.
\end{align}
For the resource allocation design of the AF relay-assisted system, we jointly optimize the transmit covariance matrices, $\bQ_k$, $\forall k$, at the UT sides and the AF matrix, $\bF$, to maximize the system GEE, which is characterized as
\begin{subequations}\label{eq:problem_Q_F}
\begin{align}
\cP_{\bQ,\bF}:\quad\underset{\bQ,\bF} \max \quad & \frac{W R_{\AF}}{P_{\AF,\tot}} \\
{\mathrm{s.t.}}\quad
& \tr { \bQ_k } \le P_{\max,k}, \quad \bQ_k \succeq \bzero,\quad\forall k \in \cK, \\ \label{eq:AF_constraint}
& P_{\AF}(\bF) \le P_{\max,\AF},
\end{align}
\end{subequations}
where $P_{\max,\AF}$ depends on the relay power budget. For fair comparison, we set $P_{\max,k} = P_{\max,\AF} = P_{\max}$, $\forall k$. In addition, the thermal noise variance, the hardware dissipated power per antenna, and the amplifier inefficiency factor at the relay station are set as $\sigma^2_{\AF} = -120$ dBm, $P_{\mathrm{s},\AF} = 10$ dBm, and $\xi_{\AF} = 1/0.3$, respectively.

The GEE maximization problem $\cP_{\bQ,\bF}$ for the relay-assisted system is also tackled via utilizing the AO method. Given an arbitrary $\bF$, we can optimize $\bQ$ by performing a similar approach as that in \secref{sec:Optimization_Q}. However, since the denominator of the objective in \eqref{eq:problem_Q_F} is related to $\bF$ and constraint \eqref{eq:AF_constraint} is challenging to handle, the optimization of $\bF$ with fixed $\bQ$ is quite different from that of $\bPhi$. Hence, numerical exhaustive search is employed to optimize $\bF$ \cite{huang2019reconfigurable}.

\begin{figure}
\centering
\includegraphics[width=0.6\textwidth]{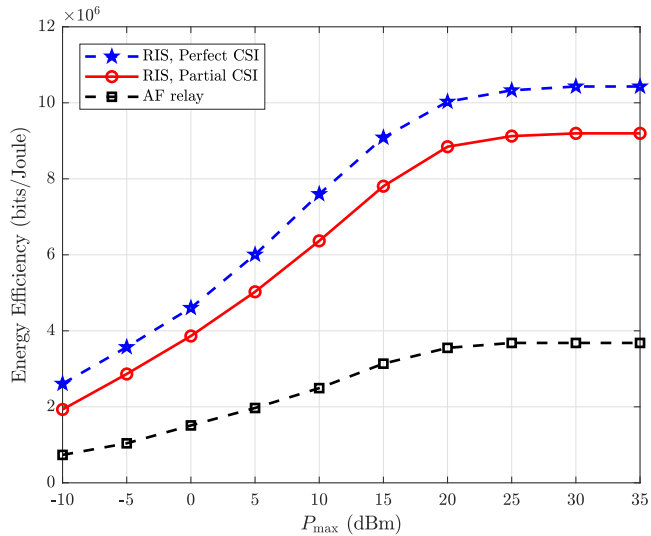}
\caption{Comparison of the GEE performance versus $\Pmax$ in RIS and AF relay-assisted systems.}
\label{fig:Comparison_AF}
\end{figure}

In addition, we consider the scheme exploiting perfect instantaneous CSI of both UT-to-RIS and RIS-to-BS channels, which serves as the comparison benchmark. \figref{fig:Comparison_AF} illustrates the comparison of the GEE performance between the proposed RIS-assisted transmission scheme with partial CSI and other schemes, including the perfect CSI benchmark case as well as the AF relay-assisted baseline case. It is shown that the RIS-assisted system significantly outperforms the AF relay-assisted one in terms of GEE. This can be explained by the fact that the relay-assisted system exhibits higher energy consumption compared to the RIS-assisted one. As mentioned before, passive RIS elements reflect received signals without adopting a transmitter module while the active AF relay assists transmissions through generating new signals, which incurs additional transmit power consumption. In addition, better GEE performance can be attained using perfect CSI but with a larger signaling overhead. Furthermore, it is interesting to notice that the gaps among the curves of the three schemes remain constant in the high transmit power budget. This behaviour is due to the reason that the system GEE will eventually saturate for large power budgets.

\section{Conclusion}\label{sec:conclusion}

We investigated resource allocation for RIS-assisted multiuser MIMO uplink communication systems under the GEE maximization criterion. The transmit covariance matrices of the UTs and the phase shifts of the RIS reflector were jointly optimized in the transmission design, subject to a transmit power constraint at each UT. We considered a practical scenario, where the instantaneous knowledge of the RIS-to-BS channel is available, while only the statistical knowledge of the UT-to-RIS channels can be exploited for resource allocation. We first obtained closed-form solutions for the optimal transmit signal directions at the UT sides. Taking advantage of the random matrix theory, we simplified the subsequent optimizations with a DE-based objective function. Then, we utilized Dinkelbach's approach to solve the power allocation problem with a fixed RIS phase shift matrix. In addition, to optimize the RIS phase shift matrix, we introduced an equivalent MSE minimization problem, which was tackled by the BCD method as well as the MM technique. Demonstrated by numerical results, the developed approach is effective in both GEE and SE maximization. Moreover, the RIS-assisted systems can achieve significant GEE performance gains compared to some traditional baseline schemes.

\appendices

\section{Proof of \propref{prop:mm_convergence}}\label{app:B}
To gain some insight into the properties of the adopted MM technique, consider the general minimization program as follows
\begin{align}
\cP : \quad \underset{\bx \in \cX } \min \quad & { f(\bx) } \ntb
{\mathrm{s.t.}}\quad
& \bx \in \cX,
\end{align}
where $\cX$ is a convex and compact feasible set. Denote by $\cP^{(i)} = \underset{\bx \in \cX } \min \ { \widetilde{f} (\bx | \bx^{(i)})}$ a series of minimization programs with $\bx^{(i+1)}$ being the corresponding minimizer. The surrogate objective function $ \widetilde{f} (\bx | \bx^{(i)})$ in each $\cP^{(i)}$, is approximate to $f(\bx)$, which is constructed by the previous optimal solution $\bx^{(i)}$. In addition, problem $\cP^{(i)}$ has the same feasible set $\cX$ as that of problem $\cP$. If the surrogate objective functions $ \widetilde{f} (\bx | \bx^{(i)})$ have the following properties:
\begin{description}
\item[$1)$] $\widetilde{f} (\bx | \bx^{(i)}) \ge f(\mathbf{x}),\ \forall \bx$,
\item[$2)$] $\widetilde{f} (\bx^{(i)} | \bx^{(i)}) = f({\bx^{(i)}})$,
\item[$3)$] $\nabla_{\bx} \ \widetilde{f} (\bx^{(i)} | \bx^{(i)}) = \nabla_{\bx} \ f({\bx^{(i)}})$,
\end{description}
we can then conclude that the minimum sequence $\left\{\widetilde f\left(\bx^{(i+1)}| \bx^{(i)}\right)\right\}_{i=0}^{\infty}$ is monotonically non-increasing and convergent. In addition, the optimizer sequence $\left\{\bx^{(i)}\right\}_{i=0}^{\infty}$ converges to a resulting point $\bx^*$ fulfilling the first-order optimality conditions of $\cP$ \cite{marks1978general}.

For the phase shift optimization problem in \eqref{eq:MSE_minimization_Phi}, the three properties described above can be readily checked to be satisfied with respect to $\widetilde{f} \left(\bphi | \bphi^{(i)}\right)$ given in \eqref{eq:g_upperbound} \cite{huang2019reconfigurable}. As a consequence, the results in \emph{\propref{prop:mm_convergence}} hold. This concludes the proof.


\end{document}